\documentclass[11pt,notitlepage]{article}

\usepackage[english]{babel}

\usepackage{xcolor}
\usepackage{hyperref}
\usepackage{IEEEtrantools}

\usepackage{titlesec}
\usepackage{titletoc}

\hypersetup{
    colorlinks=true,
    linkcolor=black,
    citecolor=black,
    urlcolor=black
}

\usepackage[letterpaper,top=2cm,bottom=2cm,left=3cm,right=3cm,marginparwidth=1in]{geometry}

\usepackage{amsmath,amssymb}
\usepackage{booktabs}
\usepackage{graphicx}
\usepackage{subcaption}
\usepackage{multirow}
\usepackage{rotating}
\usepackage{tabularx}
\usepackage{tikz}
\usepackage{fancyvrb}
\usepackage{inconsolata}
\usetikzlibrary{arrows.meta,fit,positioning}

\newcommand{\topstack}[1]{\begin{tabular}[t]{@{}l@{}}#1\end{tabular}}

\title{The AnyLog Edge Data Fabric}

\author{
Roy Shadmon, PhD \\
\and Mark Davidson \\
\and Eric Aquaronne \\
\and Massimiliano Pinto \\
\and Ori Shadmon \\ 
\and Moshe Shadmon \\
}

\date{}

\begin{document}

\maketitle

\bstctlcite{BSTcontrol}

\begin{abstract}

Industrial and autonomous systems increasingly depend on AI, automation, and real-time coordination to act on operational data as it is generated. Yet conventional architectures often require that data to pass through centralized platforms before decisions can be made. Cloud systems remain valuable for training, reporting, and long-term analytics, but they add latency and external dependencies to the critical decision path and become harder to scale as each new site adds additional edge devices and data. As intelligence spreads across machines, sites, facilities, and vehicles, continued dependence on centralization will constrain response time, resilience, scalability, and autonomous operation.

This paper presents the AnyLog Edge Data Fabric, an agent- and edge-based platform that manages operational data at its source while presenting distributed data, assets, compute resources, and services as one logical system. Through its Distributed Metadata Layer, Virtual Data Lake, Unified Namespace, Single System Image, and Model Context Protocol, authorized users, applications, automation services, and AI agents can discover, query, process, and act on distributed resources without knowing where they are hosted. Queries and computation execute at the agents holding the relevant data, so only requests and results traverse the network. This preserves local ownership, reduces data movement, supports continued operation during connectivity disruptions, and enables repeatable deployment from validated digital-twin configurations. AnyLog provides a cloud-like operating model for distributed SQL, real-time automation, Edge AI, federated learning, and resilient decision-making without a single point of failure or any dependence on centralized infrastructure.

\end{abstract}

\startcontents
\printcontents{}{0}[2]{}

\section{Introduction}

The next generation of industrial AI depends on bringing intelligence to the data rather than moving operational data to centralized platforms. As autonomous systems take greater responsibility for monitoring, coordination, and control, direct access to current conditions at the edge becomes a fundamental architectural requirement. Industrial systems cannot afford to discover critical events only after data has been transmitted, processed, and centralized. A failing manufacturing cell can stop an entire production line, instability at one substation can cascade across the grid, an autonomous vehicle can lose critical routing context when remote services are unavailable, and a deteriorating patient can cross a critical threshold while relevant data is still moving through the system. In each case, the value of an observation declines rapidly with time: information that could have prevented a failure becomes evidence explaining why it occurred.

Centralized architectures make this delay difficult to avoid because operational data must often be transmitted, ingested, normalized, and stored before applications can use it. Alerting systems, rule engines, automation services, and AI consequently operate on a delayed representation of the physical environment rather than its current state~\cite{shi2016edge,iorga2018fog}. Network disruptions, ingestion backlogs, and synchronization delays can further postpone detection and response or interrupt the decision path(s) entirely.

The problem worsens as deployments expand. Every new facility, vehicle, production line, or plant adds data that the shared network, ingestion, storage, and processing infrastructure must absorb, increasing reliance on historians, enterprise databases, data warehouses, and cloud data lakes. These systems remain essential for process history, compliance, auditing, reporting, retention, and large-scale analytics~\cite{lu2023additive,theodoratos1997data,hai2023data}, but they inherently introduce architectural risk when they are also the default path for real-time decisions, causing local operations to depend on the capacity, availability, and responsiveness of centralized infrastructure.

The architectural requirement is therefore twofold: time-sensitive processing must remain close to operational data, and authorized applications, machines, and AI agents must be able to reason across relevant observations that span across the edge rather than from isolated local views. This is essential for enabling autonomous operations like lights-out factories, where AI, analytics, rule execution, and machine-to-machine coordination operate continuously~\cite{turner2021human,liu2019edge}. The critical decision path thus requires an operational layer that makes edge data immediately available through a unified view, while forwarding selected records, key events, and summaries to centralized systems for retention, compliance, reporting, and enterprise analytics.

\begin{figure}
    \centering
    \includegraphics[width=0.9\linewidth]{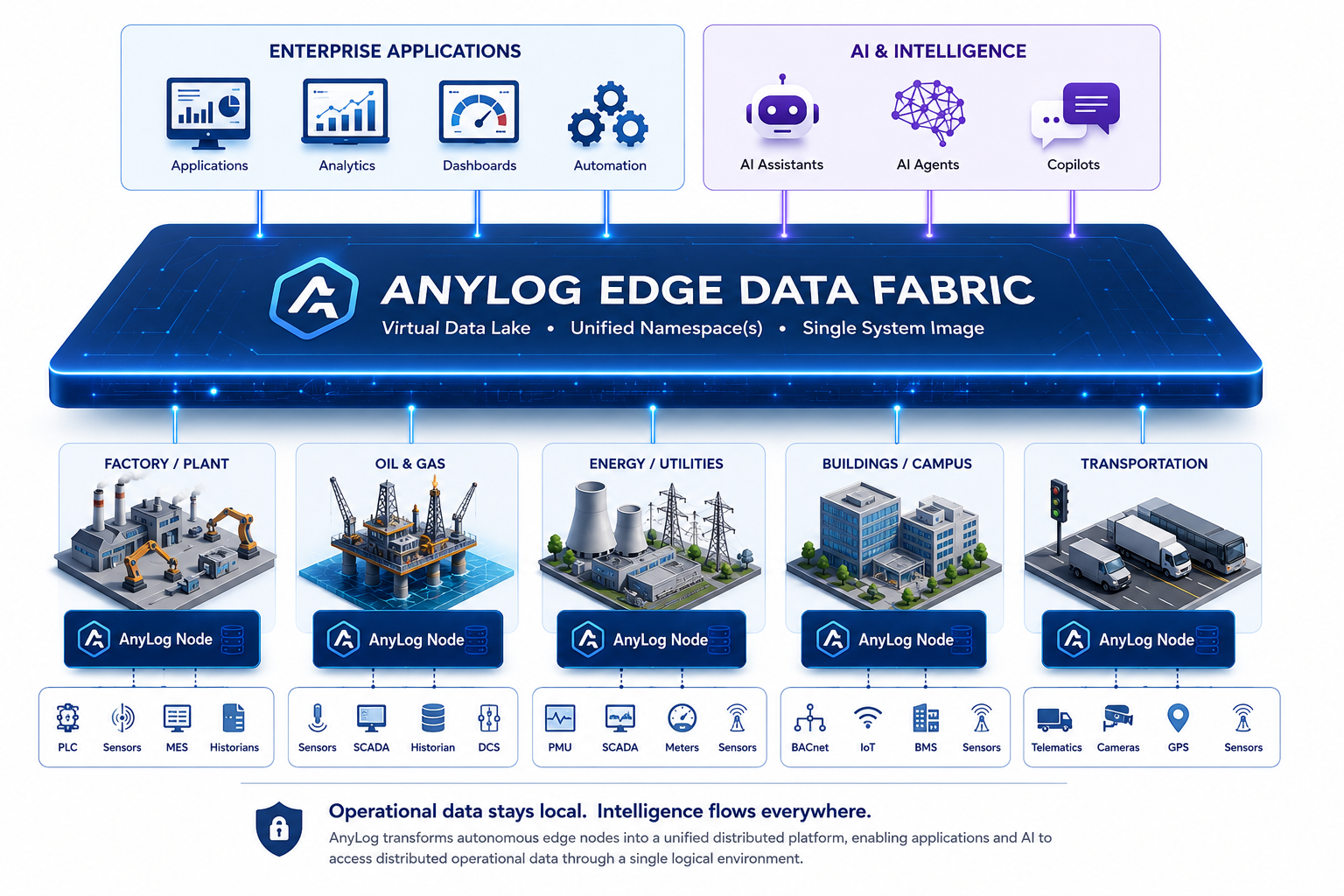}
    \caption{AnyLog EDF separates the distributed data plane from the logical control and access plane. Operational records remain in local databases at each site, while metadata, query routing, and virtualization layers provide applications, administrators, and AI with a unified view of distributed data, assets, and infrastructure.}
    \label{fig:anylog-edf}
\end{figure}

This paper presents the AnyLog Edge Data Fabric (EDF), an edge-based platform that deploys lightweight software agents across geographically distributed environments to manage operational data as one continuously connected system.
As illustrated in Figure~\ref{fig:anylog-edf}, data remains in the local databases closest to the machines, facilities, vehicles, and devices that generate it, while AnyLog presents the associated data, metadata, compute resources, services, and assets through a unified logical platform. Authorized users, applications, automation services, and AI agents can issue requests through governed interfaces without knowing where the relevant resources are physically located. AnyLog discovers the appropriate agents, coordinates distributed execution, and returns a unified result~\cite{abadi2020anylog,nawab2024tipping,shadmon2020peer}.

This cloud-like operating model is expressed through several complementary views. Applications query a \textbf{Virtual Data Lake} rather than individual databases, physical assets and their relationships are organized through one or more \textbf{Unified Namespaces (UNS)}~\cite{peter2024impact}, administrators manage the deployment through a \textbf{Single System Image}, and AI agents access the same operational context and platform services through AnyLog's \textbf{Model Context Protocol (MCP) server} hosted by each agent. 

The architecture is built on three simple principles:

\begin{enumerate}
\item Keep operational data where it is generated.
\item Move processing to the data.
\item Share metadata rather than raw data.
\end{enumerate}

The remainder of this paper explains how these principles are realized within the AnyLog Edge Data Fabric. Although the examples focus on industrial and autonomous systems, the architecture applies wherever distributed data must be discovered, processed, and acted upon in real-time without sacrificing locality, ownership, performance, or resilience.

The sections that follow develop these principles into the complete AnyLog architecture. Section~\ref{ch:architecture} describes how autonomous AnyLog agents implement the platform, and Section~\ref{ch:query} covers distributed SQL execution. Section~\ref{ch:principles} explains how the modular, distributed design integrates new data sources and sites, adapts as deployments evolve, and presents the expanding environment as one unified data platform. Section~\ref{ch:integration} describes integration with sensors, historians, databases, applications, and cloud platforms. Section~\ref{chap:operating-platform} addresses management, observability, diagnostics, and security, while Section~\ref{chap:ai-native-platform} extends the architecture to Edge AI, federated learning, and mobile autonomous systems. Section~\ref{ch:conclusion} concludes the paper.

% The sections that follow develop these principles into the complete AnyLog architecture. Section~\ref{ch:architecture} explains how autonomous AnyLog agents realize those principles. 
% Section~\ref{ch:principles} explains how the AnyLog Edge Data Fabric uses a modular, distributed architecture to integrate new data sources and sites, adapt as deployments evolve, and present the expanding environment as one unified data platform.
% Section~\ref{ch:query} presents distributed SQL execution, while Section~\ref{ch:integration} describes how sensors, historians, databases, applications, and cloud platforms integrate with the fabric. Section~\ref{chap:operating-platform} covers platform management, observability, diagnostics, and security. Section~\ref{chap:ai-native-platform} extends the architecture to Edge AI, federated learning, and mobile autonomous systems. Section~\ref{ch:conclusion} concludes the paper.

% AnyLog distributes the knowledge and computation required to locate, understand, and process operational data wherever it resides. Each agent retains and manages the data generated within its environment, while the Distributed Metadata Layer provides the shared knowledge needed to identify datasets, resolve their physical locations, understand their schemas, enforce access policies, and coordinate work across participating agents. Individually, the agents remain autonomous; together, they operate as one logical platform.

\section{The AnyLog Architecture}
\label{ch:architecture}

\subsection{The AnyLog agent}
\label{sec:roles}

The fundamental building block of the platform is the \textbf{AnyLog agent}: independently operating software that can run on industrial gateways, embedded computers, edge servers, virtual machines, cloud instances, Raspberry Pis, or enterprise servers. Every agent uses the same core software and can assume one or more operational roles, including data collection, storage, query execution, and agentic services. These roles are configured according to the compute, memory, storage, and network resources available on each device, allowing a single software stack to support deployments ranging from lightweight data collection to fully integrated data, query, and AI systems.

An \textbf{Operator} stores operational data in locally managed databases, enforces the policies governing access to that data, and makes authorized records available for distributed queries and services. Operators execute queries over their local data, apply filtering and aggregation, and return raw records, query results, or key events. By managing data near the assets they represent, each Operator contributes a subset of the data available through the fabric's unified logical view.

A \textbf{Publisher} is used when the system collecting data from PLCs or sensors lacks the compute or storage resources to retain and manage that data locally. The Publisher captures raw data near its source and forwards it to an Operator or another configured destination.

A \textbf{Query Coordinator} acts as the gateway to the Edge Data Fabric. It receives requests from users, applications, dashboards, automation services, or AI agents; uses metadata to locate the Operators that host the relevant data and services; coordinates execution across participating Operators; and returns a unified result. 

An \textbf{MCP Provider} serves as a Query Coordinator for authorized AI agents, exposing the Edge Data Fabric through a local Model Context Protocol server~\cite{mcpSpecification}. It provides governed access to distributed data, tools, schemas, asset relationships, and platform state without requiring the AI agent to discover or manage the AnyLog agents hosting those resources.

A \textbf{Metadata Manager} distributes the shared metadata required for discovery, policy enforcement, and coordination. Metadata Managers synchronize information about schemas, data locations, agent capabilities, access policies, services, and Unified Namespace relationships so that agents can locate resources and cooperate without point-to-point configuration.

In deployments that require a more decentralized coordination model, metadata can also be alternatively managed through a blockchain-backed layer. Programmable smart contracts allow agents to publish, synchronize, and receive metadata updates through distributed mechanisms such as publish-subscribe~\cite{khan2021blockchain,zheng2024decentagram,cai2022rbaas}.

\subsection{The Distributed Metadata and Policy Layer}
\label{sec:metadata}

Cooperation among AnyLog agents depends on a shared understanding of the distributed environment. The \textbf{Distributed Metadata Layer} provides this common view by describing what data, assets, services, and agents exist, where they reside, how they relate, and how they may be accessed. Because metadata is far smaller than the operational data itself (Section~\ref{sec:coordinate-to-reduce-network}), it can be synchronized efficiently across the Edge Data Fabric, giving each agent the context needed to discover and use distributed resources.

Policies translate this shared context into enforceable behavior. They define how agents are configured and how data and services are accessed, processed, retained, shared, and protected. Enforcement remains local to the agents responsible for each resource, preserving consistent governance without introducing a central control dependency.

\textbf{Node Policies} enable repeatable, one-click deployment by defining how agents configure themselves for one or more operational roles, as described in Section~\ref{sec:config}. A policy may reference deployment scripts, database settings, services, security requirements, and other role-specific configuration. The configuration is defined once, reused across multiple agents, and automatically reapplied when an agent joins or restarts. Multiple policies may exist for the same role, allowing each Operator, Publisher, Query Coordinator, or other agent to follow the configuration assigned to its environment.

\textbf{Access Policies} define which identities may access specific services, databases, tables, and time ranges. These policies are distributed through the metadata layer and enforced locally by the agent that owns the requested data or service. If no applicable policy authorizes the request, access is denied.

\textbf{Table Policies} define a logical table within the Edge Data Fabric. They describe its schema, the Operators that host the data, the physical database and table mappings, and the metadata required for AnyLog to discover the data and execute distributed queries through a single logical table.

\textbf{Unified Namespace Policies} define the hierarchical relationships among assets. Each policy identifies a root asset and maps its parent-to-child structure, such as an organization containing facilities, a facility containing production lines, a line containing machines, and a machine containing sensors. These relationships create consistent logical paths that enable users, applications, and AI agents to discover and access the data, services, and metadata associated with any asset or branch of the hierarchy without requiring the underlying data to be physically reorganized or centralized in systems such as MQTT brokers, historians, or cloud platforms.

\textbf{Rule Engine Policies} define how agents evaluate incoming events, execute scheduled actions, and respond to operational conditions without waiting for a centralized application. A policy may specify that if a PLC fault bit changes from 0 to 1, torque on a fastening cell exceeds its accepted range for five consecutive cycles, or temperature moves outside a permitted band, the agent should write an alert row, publish a Unified Namespace update, send an SMS or email notification, forward the event to a historian, or invoke a downstream workflow. Because these rules execute locally, time-sensitive responses continue even when wide-area connectivity is slow, intermittent, or unavailable.

\textbf{Aggregation Policies} define how high-volume observations are reduced into useful summaries. If 10{,}000 consecutive inputs report the same value, an agent may store one row containing the start timestamp, end timestamp, received-record count, and observed value rather than retaining 10{,}000 identical records. The same principle can be applied to tolerance bands, allowing readings that remain within \(\pm \epsilon\) to be represented by their time span, count, representative value, and observed bounds. Policies may also define time-windowed aggregates, such as storing the count, sum, minimum, maximum, median, and average of a measurement stream every 10 minutes. This reduces storage and network demand by enabling pre-processing directly at the edge.

\textbf{Retention and Forwarding Policies} define how long different classes of data remain at the edge and which information is forwarded to enterprise systems. A deployment may retain detailed raw records locally for 10 days while sending selected summaries, regulated records, significant events, alarms, and alerts to a historian or cloud repository through native northbound connectors. As data ages, policies may archive it to object storage, file systems, cloud repositories, or other approved storage systems. The schemas and archive locations remain represented in the Distributed Metadata Layer so that historical information stays discoverable across storage tiers.

\textbf{Replication Policies} preserve access to selected datasets, summaries, and event streams when an Operator becomes unavailable. According to policy, the specified data is replicated to designated agents, and each replica is represented in the table metadata under the same cluster identifier as the primary. During query planning, if the primary agent is unavailable, the Distributed Query Engine routes the request to an available replica within that cluster, as described in Sections~\ref{sec:system-model} and~\ref{sec:virtualization}.

Together, the Distributed Metadata Layer and its associated policies allow autonomous agents to operate as one coordinated platform. The metadata establishes a shared view of the environment, while locally enforced policies create a single system image view across data management, automation, security, resilience, query execution, and AI access.

\subsection{System Model}
\label{sec:system-model}

Let \(\mathcal{N} = \{n_1,\dots,n_k\}\) denote the set of AnyLog agents participating in a deployment. Each agent \(n \in \mathcal{N}\) maintains a local policy set \(P_n\), a local metadata view \(M_n\) synchronized through the Distributed Metadata Layer, and a set \(D_n\) of local data stores, which may include SQL databases, NoSQL stores, and object or blob stores such as S3-compatible buckets, and which may be empty for agents that do not store operational records.

Let
\[
\mathcal{T} = \{\mathsf{pub}, \mathsf{op}, \mathsf{meta}, \mathsf{qry}\}
\]
denote the set of agent types, corresponding respectively to publisher (pub), operator (op), metadata-manager (meta), and query (qry) roles. Each agent is assigned a role profile
\[
\tau : \mathcal{N} \rightarrow 2^{\mathcal{T}},
\]
where \(\tau(n)\) may contain one or more roles because the same AnyLog runtime can be configured to combine them.

The roles are interpreted as follows.
\begin{itemize}
    \item If \(\mathsf{pub} \in \tau(n)\), agent \(n\) can collect or receive source data and publish it to one or more operator agents. This role is useful when the machine closest to the data source is resource-constrained and cannot be utilized for storage or query processing.
    \item If \(\mathsf{op} \in \tau(n)\), agent \(n\) operates as an Operator: it stores operational data in one or more local data stores, including SQL databases, NoSQL systems, and object or S3-like blob stores. It executes the supported retrieval, filtering, aggregation, or query operations over its local data and returns records or partial results for distributed execution.
    \item If \(\mathsf{meta} \in \tau(n)\), agent \(n\) acts as a metadata manager in deployments that do not use blockchain-backed metadata synchronization. In that mode, metadata-manager agents disseminate metadata updates and answer metadata queries for discovery and coordination.
    \item If \(\mathsf{qry} \in \tau(n)\), agent \(n\) can accept client requests, resolve metadata, coordinate distributed queries, and return unified results.
\end{itemize}

Define the operator-agent subset
\[
\mathcal{N}_{\mathsf{op}} = \{ n \in \mathcal{N} \mid \mathsf{op} \in \tau(n)\}.
\]
Only operator agents host local table fragments. Publisher agents feed data into operator agents, query coordinators coordinate requests, and metadata-manager agents are needed only in manager-based metadata deployments.

Metadata dissemination follows one of two deployment modes. In blockchain-backed deployments, agents synchronize against the shared blockchain ledger-backed metadata state and the metadata-manager role may be omitted. In manager-based deployments, let
\[
\mathcal{N}_{\mathsf{meta}} = \{ n \in \mathcal{N} \mid \mathsf{meta} \in \tau(n)\}
\]
denote the metadata-manager subset responsible for propagating and serving metadata.

The application-facing logical data surface is a set \(\mathcal{V}\) of \textbf{virtual tables}. Each virtual table is identified by a tuple
\[
v = (c,d,t) \in \mathcal{V}
\]
where \(c\) is a logical grouping, \(d\) is a logical DBMS name, and \(t\) is a logical table name.

Physical placement is described by a metadata mapping
\[
\mu : \mathcal{V} \rightarrow 2^{\mathcal{N}_{\mathsf{op}} \times \mathcal{R} \times \mathcal{K}}
\]
where \(\mathcal{R}\) is the set of serving roles, such as primary, backup, or partition member, and \(\mathcal{K}\) is the set of cluster identifiers. If \((n,r,k) \in \mu(v)\), then agent \(n\) can serve virtual table \(v\) in role \(r\) within cluster \(k\).

For each agent \(n\) and virtual table \(v\), let \(F_n(v)\) denote the local fragment of \(v\) stored on \(n\). By default, AnyLog treats virtual tables as \textbf{partitioned}: different agents generally hold different fragments, and the logical table is the union of those fragments across the eligible serving agents. Let
\[
\mathcal{N}(v) = \{ n \in \mathcal{N}_{\mathsf{op}} \mid \exists r \in \mathcal{R}, k \in \mathcal{K} : (n,r,k) \in \mu(v)\}.
\]
Then the logical content of \(v\) is
\[
v \equiv \bigcup_{n \in \mathcal{N}(v)} F_n(v).
\]
When a replication policy \(\rho_v\) is enabled, selected fragments may also be copied periodically to designated backup agents. Replication therefore improves availability, but does not change the logical definition of \(v\). Query planning selects one eligible copy of each replicated fragment so that replicated storage does not duplicate logical query results.

A distributed query \(q\) over a virtual table \(v\) is evaluated by selecting an eligible participant set \(S(q,v) \subseteq \mathcal{N}(v)\) from metadata, local policies, and replica eligibility. Each participating agent evaluates \(q\) on its local fragment and returns either records or partial aggregates. The coordinator then computes
\[
\operatorname{Ans}(q,v) =
\operatorname{Combine}\left(\left\{
\operatorname{Eval}(q,F_n(v)) \mid n \in S(q,v)
\right\}\right).
\]
This captures the main semantics of the Virtual Data Lake: applications see one logical table even though storage, policy enforcement, and most execution remain distributed.

\subsection{Virtualizing the Distributed Environment}
\label{sec:virtualization}

\begin{figure}
    \centering
    \includegraphics[width=0.92\linewidth]{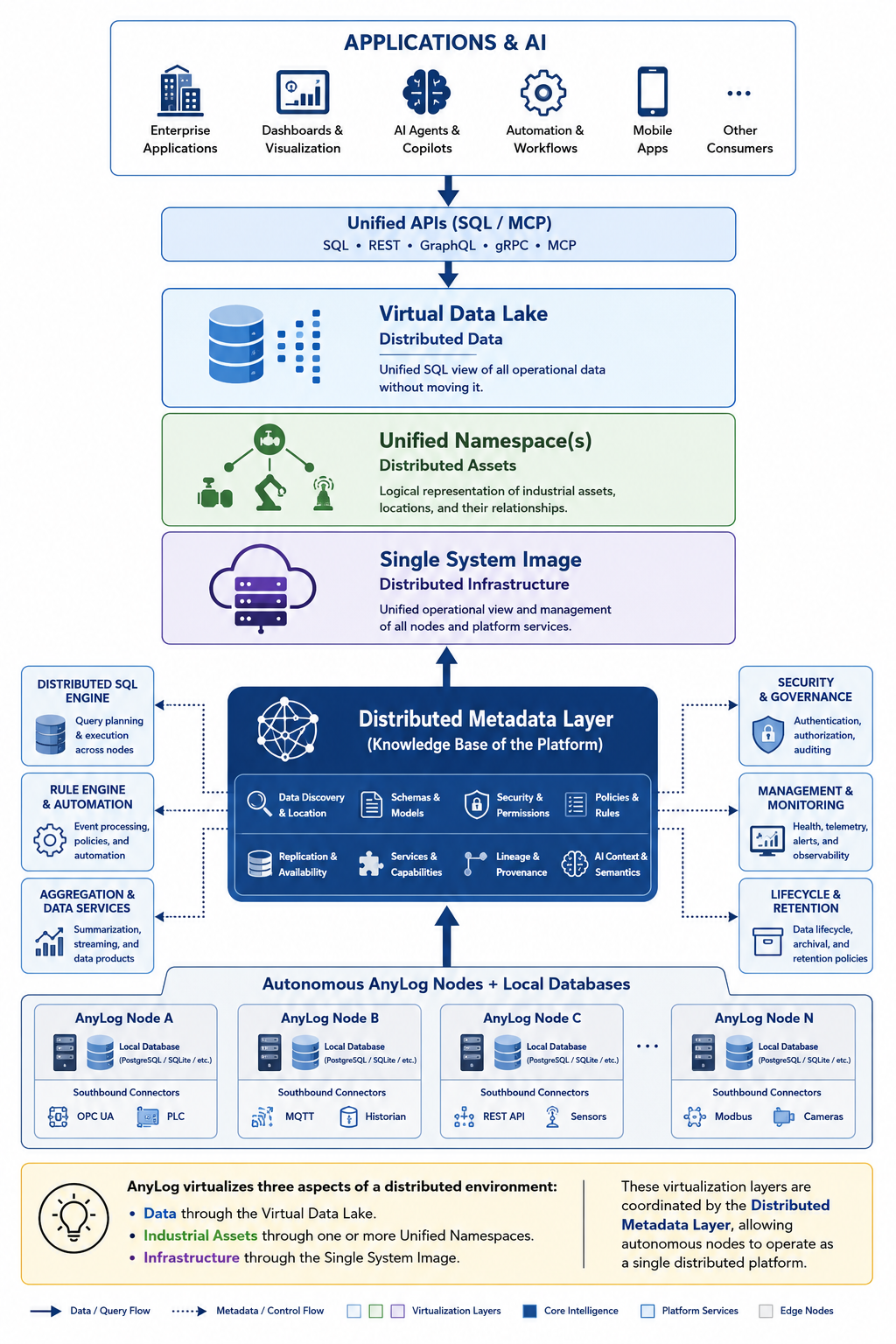}
    \caption{Architecture overview of AnyLog EDF. Autonomous agents keep operational data local, publish metadata to the distributed control plane, and expose their resources through three logical views: a Virtual Data Lake for data access, one or more Unified Namespaces for asset context, and a Single System Image for platform operations.}
    \label{fig:architecture}
\end{figure}

The purpose of the AnyLog architecture is to abstract the complexity of distributed edge infrastructure by presenting it as a Single System Image (SSI), allowing users, applications, and AI agents to discover, access, and manage distributed resources through a unified logical interface. As illustrated in Figure~\ref{fig:architecture}, the three virtualization layers work together to create this user-facing view, with the Layer~3 SSI serving as the final interface through which the underlying distributed environment is accessed and managed.

The \textbf{Virtual Data Lake} presents distributed operational edge data as though it were contained within a single logical database. Applications query distributed data using standard SQL without needing to know where data is physically stored, described in Section~\ref{ch:query}.

The \textbf{Unified Namespace} provides a logical representation of industrial assets independent of the databases that contain their operational information. Applications and AI navigate logical assets and their relationships, while the mapping layer resolves the underlying physical systems and relational tables. Multiple namespaces can coexist to support different operational perspectives over the same distributed data.

The \textbf{Single System Image} virtualizes the infrastructure itself. Administrators monitor and manage geographically distributed agents through a unified operational view that resembles the experience of managing a centralized cloud platform. Together, these three virtualization layers transform a distributed collection of autonomous systems into a cloud-like operational environment where operational data remains local, processing executes at the edge, and every agent retains its autonomy.

\begin{table}[t]
    \centering
    \footnotesize
    \setlength{\tabcolsep}{6pt}

    \begin{tabular}{
        @{}
        l
        l
        l
        @{}
    }
        \toprule
        \textbf{Company} &
        \textbf{DBMS} &
        \textbf{Logical Table} \\
        \midrule

        \multirow[t]{3}{*}{Smart City}
        & \multirow[t]{2}{*}{\texttt{cos}}
        & \texttt{wp\_analog} \\
        &
        & \texttt{pp\_pm} \\
        &
        \texttt{monitoring}
        & \texttt{agent\_insight} \\

        \addlinespace
        \multirow[t]{2}{*}{AnotherPeak}
        & \multirow[t]{2}{*}{\texttt{battery}}
        & \texttt{pack\_logs} \\
        &
        & \texttt{pack\_current} \\

        \addlinespace
        CarPlant
        & \texttt{assembly}
        & \texttt{fastening} \\

        \addlinespace
        \multirow[t]{2}{*}{Eon}
        & \multirow[t]{2}{*}{\texttt{wind\_turbine}}
        & \texttt{blade\_pitch} \\
        &
        & \texttt{power\_output} \\

        \bottomrule
    \end{tabular}

    \caption{Representative \texttt{get virtual tables} output from the AnyLog test network. This view exposes only the logical surface visible to applications and administrators: \texttt{Company}, \texttt{DBMS}, and \texttt{Logical Table}. Physical placement, partitioning, and replica membership are revealed separately through \texttt{get data nodes}.}
    \label{tab:get-virtual-tables}
\end{table}

At the logical layer, the AnyLog operational command \texttt{get virtual tables} lists the DBMSs and tables available for distributed SQL queries, organized by their logical grouping, as shown in Table~\ref{tab:get-virtual-tables}. The command presents the application-facing view of the queryable SQL tables in the fabric.

\begin{sidewaystable}[p]
    \centering
    \footnotesize
    \renewcommand{\arraystretch}{1.15}
    \setlength{\tabcolsep}{4pt}

    \begin{tabularx}{\textheight}{
        @{}
        p{1.85cm}
        p{1.9cm}
        p{2.35cm}
        p{1.25cm}
        p{4.65cm}
        >{\raggedright\arraybackslash}X
        @{}
    }
        \toprule
        Company
        & DBMS
        & Logical Table
        & \shortstack[l]{Cluster\\ID}
        & \shortstack[l]{AnyLog agent\\Endpoints}
        & Cluster Interpretation
        \\
        \midrule

        \multirow[t]{5}{1.85cm}{Smart City}
        & \multirow[t]{2}{1.9cm}{\texttt{cos}}
        & \texttt{wp\_analog}
        & \texttt{64999c}
        & \topstack{
            \texttt{P 203.0.113.41:32148}\\
            \texttt{B 198.51.100.44:32148}
        }
        & Replicated water-plant partition. Because both endpoints share cluster
          \texttt{64999c}, they are backup members of the same physical partition.
        \\

        &
        &
        \texttt{pp\_pm}
        & \texttt{2a5f83}
        & \topstack{
            \texttt{P 192.0.2.14:32148}\\
            \texttt{B 198.51.100.37:32148}
        }
        & Replicated power-plant partition. The shared cluster identifier means
          that the primary and backup members serve the same physical partition.
        \\

        &
        \multirow[t]{3}{1.9cm}{\texttt{monitoring}}
        & \multirow[t]{3}{2.35cm}{\texttt{agent\_insight}}
        & \texttt{2a5f83}
        & \topstack{
            \texttt{P 192.0.2.14:32148}\\
            \texttt{B 198.51.100.37:32148}
        }
        & Power-plant monitoring partition, replicated within one cluster.
        \\

        &
        &
        &
        \texttt{64999c}
        & \topstack{
            \texttt{P 203.0.113.58:32148}\\
            \texttt{B 192.0.2.63:32148}
        }
        & Water-plant monitoring partition, also replicated within one cluster.
        \\

        &
        &
        &
        \texttt{22be8c}
        & \texttt{P 198.51.100.89:32148}
        & Wastewater monitoring partition with one serving member. Different
          cluster IDs under the same logical table indicate independent partitions.
        \\
        \midrule

        \multirow[t]{4}{1.85cm}{AnotherPeak}
        & \multirow[t]{4}{1.9cm}{\texttt{battery}}
        & \multirow[t]{2}{2.35cm}{\texttt{pack\_logs}}
        & \texttt{fec25e}
        & \texttt{P 203.0.113.103:32148}
        & Vessel partition A.
        \\

        &
        &
        &
        \texttt{ec5697}
        & \texttt{P 198.51.100.122:32148}
        & Vessel partition B. The logical table spans two clusters, so the data
          is partitioned across independent clusters.
        \\

        &
        &
        \multirow[t]{2}{2.35cm}{\texttt{pack\_current}}
        & \texttt{ec5697}
        & \texttt{P 198.51.100.122:32148}
        & Derived current-data partition A.
        \\

        &
        &
        &
        \texttt{fec25e}
        & \texttt{P 203.0.113.103:32148}
        & Derived current-data partition B. The table follows the same
          two-cluster distribution pattern as \texttt{pack\_logs}.
        \\
        \midrule

        \multirow[t]{4}{1.85cm}{CarPlant}
        & \multirow[t]{4}{1.9cm}{\texttt{assembly}}
        & \multirow[t]{4}{2.35cm}{\texttt{fastening}}
        & \texttt{353495}
        & \texttt{P 192.0.2.131:32148}
        & Manufacturing-cell partition 1.
        \\

        &
        &
        &
        \texttt{0b8b65}
        & \texttt{P 203.0.113.147:32148}
        & Manufacturing-cell partition 2.
        \\

        &
        &
        &
        \texttt{65614e}
        & \texttt{P 198.51.100.166:32148}
        & Manufacturing-cell partition 3.
        \\

        &
        &
        &
        \texttt{4d1c94}
        & \texttt{P 192.0.2.184:32148}
        & Manufacturing-cell partition 4. Four independent assembly cells
          contribute records to one plant-wide logical table.
        \\
        \midrule

        \multirow[t]{8}{1.85cm}{Eon}
        & \multirow[t]{8}{1.9cm}{
            \topstack{
                \texttt{wind\_}\\
                \texttt{turbine}
            }
        }
        & \multirow[t]{4}{2.35cm}{\texttt{blade\_pitch}}
        & \texttt{4b07f4}
        & \texttt{P 198.51.100.214:32148}
        & Turbine partition 1.
        \\

        &
        &
        &
        \texttt{9366aa}
        & \texttt{P 203.0.113.201:32148}
        & Turbine partition 2.
        \\

        &
        &
        &
        \texttt{f6294d}
        & \texttt{P 192.0.2.227:32148}
        & Turbine partition 3.
        \\

        &
        &
        &
        \texttt{14866e}
        & \texttt{P 203.0.113.239:32148}
        & Turbine partition 4. The same logical table is distributed across four
          independent turbine clusters.
        \\

        &
        &
        \multirow[t]{4}{2.35cm}{\texttt{power\_output}}
        & \texttt{9366aa}
        & \texttt{P 203.0.113.201:32148}
        & Output partition 1.
        \\

        &
        &
        &
        \texttt{4b07f4}
        & \texttt{P 198.51.100.214:32148}
        & Output partition 2.
        \\

        &
        &
        &
        \texttt{f6294d}
        & \texttt{P 192.0.2.227:32148}
        & Output partition 3.
        \\

        &
        &
        &
        \texttt{14866e}
        & \texttt{P 203.0.113.239:32148}
        & Output partition 4. Multiple logical tables can share the same cluster
          topology while exposing different operational metrics.
        \\
        \bottomrule
    \end{tabularx}

    \caption{Landscape summary of representative \texttt{get data nodes}
    entries from the AnyLog test network. \texttt{Company}, \texttt{DBMS}, and
    \texttt{Logical Table} define the logical data surface, while the shortened
    \texttt{Cluster ID} and endpoint columns expose the physical AnyLog serving
    pattern. Endpoints sharing a cluster ID are backup or replica members of the
    same physical partition; different cluster IDs under the same logical table
    are independent partitions.}

    \label{tab:get-data-agents}
\end{sidewaystable}

The AnyLog command \texttt{get data nodes} exposes the physical topology behind each virtual table, as shown in Table~\ref{tab:get-data-agents}. Applications continue to address resources through the logical tuple of \textbf{Company}, \textbf{DBMS}, and \textbf{Table}, while AnyLog maps each request to the clusters and agents currently serving that resource. Agents sharing a cluster identifier represent primary ($P$) or backup ($B$) are members of the same physical partition, whereas different cluster identifiers under one logical table represent independent data fragments that AnyLog unifies under the same DBMS and table name. If a primary agent becomes unavailable, the Distributed Query Engine routes the request to a backup; if no backup exists, the query summary identifies the unavailable agent and tags the result as incomplete.

In the examples above, Smart City combines replicated and single-member partitions, AnotherPeak distributes vessel data across two clusters, CarPlant presents four manufacturing cells through one logical table, and Eon exposes tables spanning four clusters of remote wind turbines. The \texttt{Company} field adds multi-tenant organization and isolation to this logical model. For example, \texttt{CompanyA} may operate a production facility while \texttt{CompanyB} and \texttt{CompanyC} represent equipment and component suppliers. Each tenant discovers and queries only its authorized resources, while \emph{Access Policies} permit approved tables or selected records to be shared across company boundaries.

Section~\ref{sec:auto-schema} explains how schemas for newly observed tables are generated and reused. The core virtualization concept is that once a logical table is defined, the same table can be queried as one logical resource even when the rows are partitioned across many agents. AnyLog therefore allows tables to be distributed at the level that best reflects operational requirements, including by plant, production line, manufacturing cell, asset class, supplier domain, or geographic region.

\section{Distributed Query Processing}
\label{ch:query}

\begin{figure}
    \centering
    \includegraphics[width=\linewidth]{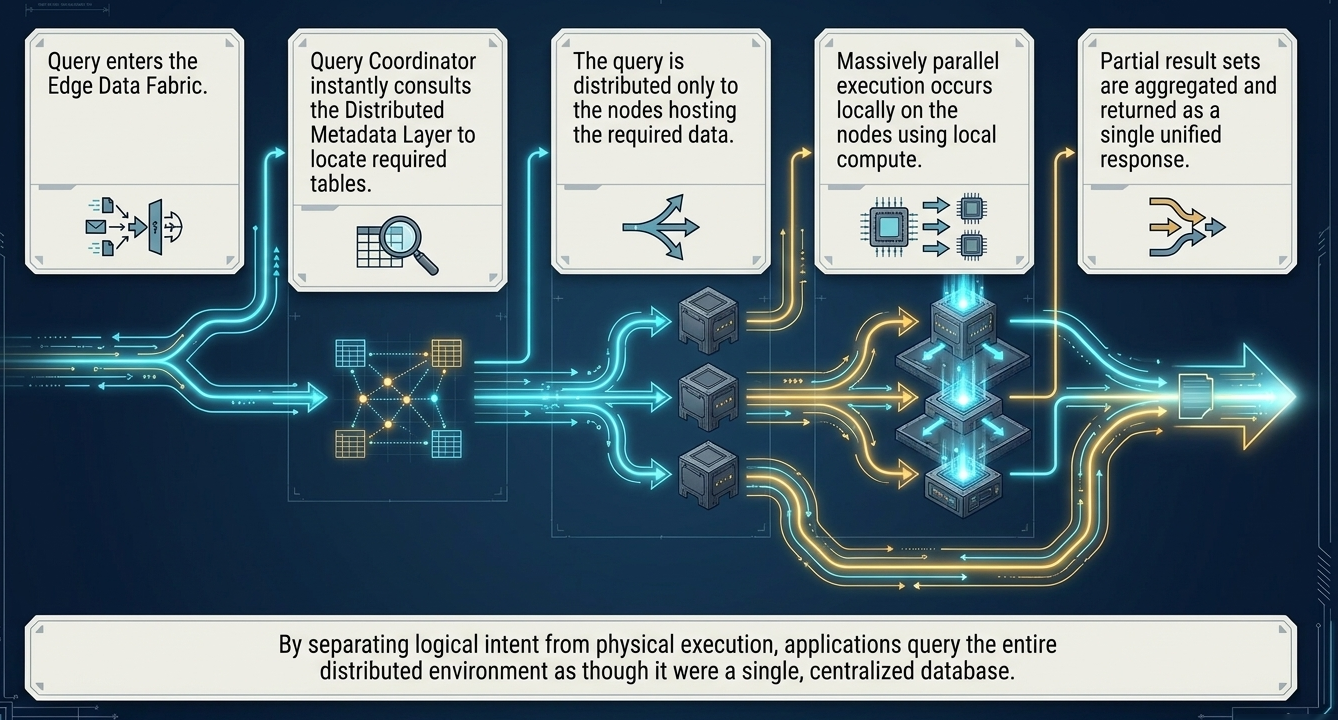}
    \caption{Five-step distributed query-processing workflow.}
    \label{fig:query-processing}
\end{figure}

\subsection{Distributed Querying Across In-Place Data}

For decades, enterprise data platforms have followed the same basic pattern: operational data is moved into a centralized repository before applications can query, analyze, or act on it. Co-locating data with a single query engine simplifies processing, but the model becomes increasingly costly and difficult to scale as industrial operations grow more distributed. High-volume data must be continuously transmitted from remote sites, shared ingestion and storage infrastructure must expand to absorb that growth, and applications may act on delayed replicas while pipelines, transformations, and synchronization complete.

\textbf{AnyLog queries operational data in place by moving computation to the agents that host the data, rather than transferring the data to a centralized processing platform.} When an application submits a request, AnyLog uses the Distributed Metadata Layer (Section~\ref{sec:metadata}) to identify the agents containing relevant data, executes the query locally and in parallel across those agents, and returns only the records, aggregates, or derived results needed to satisfy the request. The underlying operational data remains in the local databases managed by independent AnyLog agents.

From the application's perspective, the distributed environment behaves as a single logical database. The same query model applies whether the requested data resides on one agent, across several facilities, or throughout thousands of geographically distributed systems. AnyLog resolves relevant data locations, coordinates execution, enforces applicable access policies, and combines partial results into one unified response.

\subsection{A Distributed Query Engine}

Applications submit standard SQL through AnyLog's command interface, REST APIs, or Model Context Protocol (MCP) server. The command-line form is:

\begin{verbatim}
run client () sql [dbms]
SELECT [columns | aggregation]
FROM [table]
WHERE [clause]
\end{verbatim}

The target list in parentheses determines how the request is routed. An empty target list, \texttt{()}, indicates that the requester does not know which agents host the required data. AnyLog resolves the logical DBMS and table through the Distributed Metadata Layer, identifies the authorized agents that have declared relevant data, and routes the query accordingly. When specific destinations are known or required, they may be supplied explicitly:

\begin{verbatim}
run client ([ip1:port1], [ip2:port2], ...)
sql [dbms]
SELECT [columns | aggregation]
FROM [table]
WHERE [clause]
\end{verbatim}

In this form, the request is sent only to the listed agents. The same SQL statement can therefore be executed through automatic data-location discovery or against an explicitly selected set of nodes.

As illustrated in Figure~\ref{fig:query-processing}, the AnyLog agent receiving the request temporarily assumes the role of \textbf{Query Coordinator} (Section~\ref{sec:roles}). When automatic discovery is used, the coordinator consults the Distributed Metadata Layer to determine the relevant \emph{Operators} that host data associated with the query. When destinations are specified explicitly, those agents define the initial execution scope, subject to local authorization and data availability.

The Query Coordinator builds an execution plan and distributes the SQL query to the relevant agents, which execute filtering, aggregation, and retrieval locally and in parallel using their own resources. Each agent returns only the required records or partial aggregates, such as counts, sums, minima, or maxima. The Query Coordinator combines these results, applies any remaining global operations, and returns a single unified response.

Notably throughout this process, the underlying operational datasets remain in their local databases. Only query instructions and compact result sets traverse the network, reducing unnecessary data movement while allowing distributed information to be queried through one logical interface.

\subsection{Distributed Real-Time Video AI Analytics}
\label{sec:video-ai-analytics}

\begin{figure}[t]

\centering
\includegraphics[width=\linewidth]{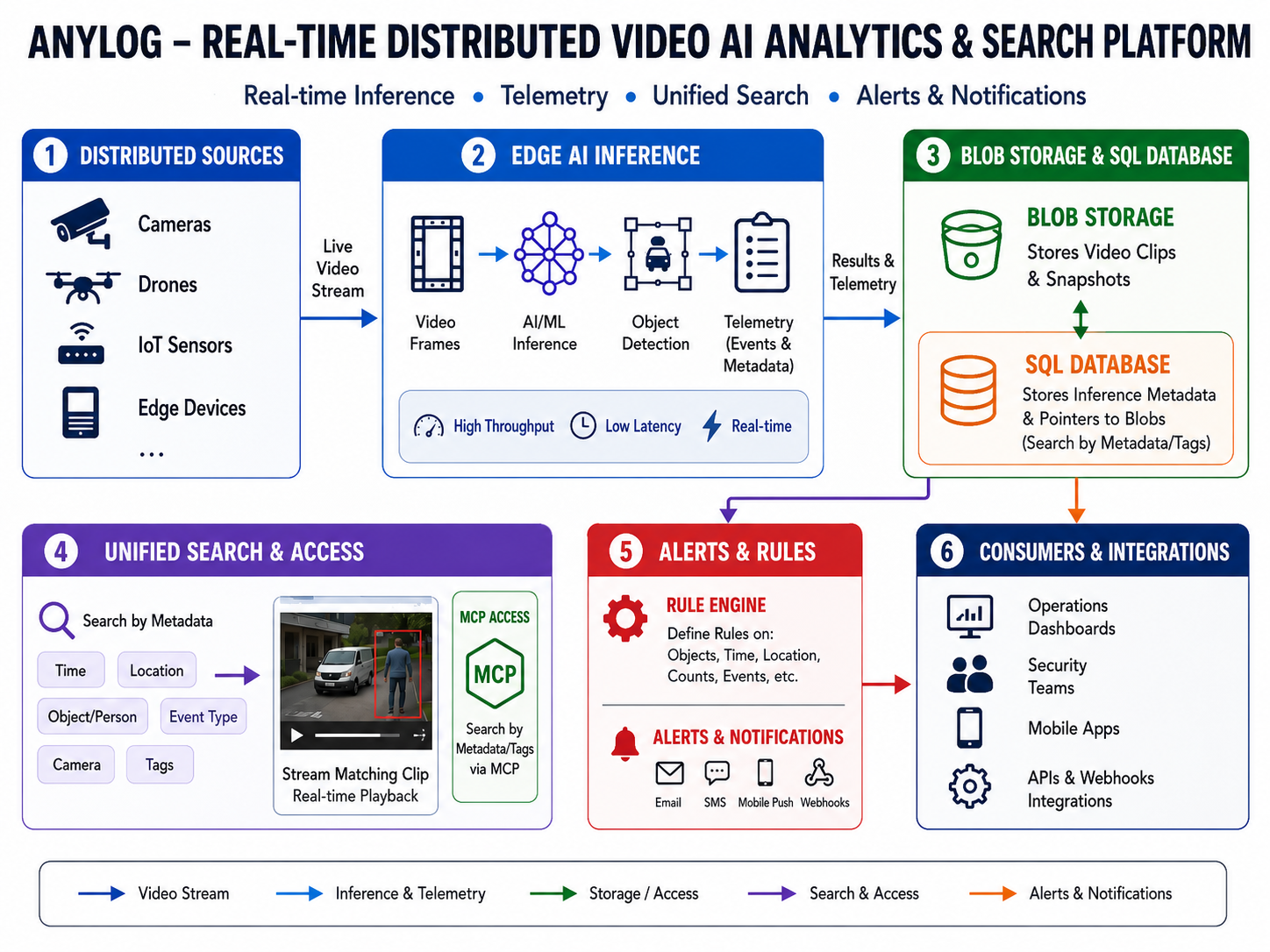}
\caption{AnyLog video architecture extending distributed telemetry capture to support real-time edge inference, with inference results and telemetry stored in local SQL databases, associated video retained in local blob storage, and unified access across independently operating edge devices.}
\label{fig:video-analytics}

\end{figure}

AnyLog extends the Single System Image beyond relational databases to include locally managed object storage, such as MinIO~\cite{minio-object-storage} and MongoDB~\cite{mongodb-gridfs}. Video segments and inference metadata remain distributed in local blob stores and SQL databases across edge devices, while the Distributed Query Engine provides searchability through one logical query interface.\footnote{Blobs may also be stored in cloud services such as Amazon S3~\cite{aws-s3}. AnyLog abstracts each storage location as another participating edge, allowing a single SQL query to locate matching video segments any blob storage location without requiring applications implement custom APIs and middleware.}

At each edge device, the AnyLog agent captures video through Southbound Interfaces such as RTSP (Section~\ref{sec:southbound}). AI models running on the device or nearby infrastructure access the live stream through Northbound Interfaces such as gRPC (Section~\ref{sec:northbound}) and perform inference on each incoming frame. Object classifications, confidence scores, timestamps, camera identifiers, locations, and user-defined tags are stored as structured telemetry in a local SQL database. The corresponding video segments remain in local blob storage, with each record referencing its supporting footage. As illustrated in Figure~\ref{fig:video-analytics}, this makes otherwise unstructured video searchable despite the video and metadata remaining at each edge.

For example, the AnyLog query below searches video inference telemetry across the logical table \nolinkurl{video_inference_events} and returns the associated blob reference, time range, detected object class, count, confidence score, and tags. The \texttt{extend} clause adds the site, camera, and hosting-agent metadata for each result, while the \texttt{selection} clause maps each record to its referenced video segment hosted in the logical table \nolinkurl{blob_reference} so the supporting footage can be retrieved directly from the physical storage location. Notably, the video segments and inference telemetry is fragmented across any number of edge devices; AnyLog builds the the query plan and fully automates the distributed query execution:
\begin{Verbatim}[commandchars=\\\{\}]
\textbf{sql} video_ai \textbf{info}=(dest_type=rest)
and \textbf{extend}=(+site_name, +camera_id, @ip, @port, @dbms_name, @table_name)
\textbf{SELECT} blob_reference, segment_start_ts::ljust(19), segment_end_ts::ljust(19), 
object_class, object_count, confidence_score, inference_tags
\textbf{FROM} video_inference_events
\textbf{WHERE} insert_timestamp >= NOW() - 20 minutes
\textbf{ORDER BY} object_count \textbf{DESC}, confidence_score \textbf{DESC}
--> \textbf{selection} (columns: ip using ip and port using port
and dbms using dbms_name and table using table_name and file using blob_reference)
\end{Verbatim}

A conventional cloud-centric architecture sends camera streams across the wide-area network (WAN) to centralized ingestion, storage, and analytics services. Detections and alerts typically become available only after the video is processed and the resulting telemetry reaches a centralized cloud database, while custom software must correlate those records with the stored blobs and video segments~\cite{aws-kinesis-video,aws-rekognition-streaming}. 
By managing video streaming and inference at the edge, the amount of data transmitted across the WAN is limited to only the segments or blobs the application or agent needs. The architecture also scales horizontally, as described in Section~\ref{sec:horizontal-scale}: each added camera, inference server, or edge device contributes capacity to the same logical environment without requiring applications to manage each resource independently.

\section{Building a Unified Foundation for Edge Intelligence}
\label{ch:principles}

\subsection{Process Data Where It Resides}

Keeping data local determines where operational records reside; processing data where it resides determines how those records are used without first centralizing them. When a request spans multiple locations, AnyLog decomposes the request and directs computation to the AnyLog agents that host the relevant data, described in Section~\ref{ch:query}. 
Each participating agent uses its local processing resources to query,  filter, aggregate, analyze, or apply AI inference to the data, returning only the information required to construct the final result. The query node then aggregates these partial results and presents them as a unified, complete result to the requesting application or AI agent.

This execution model keeps raw operational records in place. Query instructions and the metadata needed to identify relevant participants are sent to the appropriate agents, while compact records, partial aggregates, or derived results are returned to the requester. Network utilization therefore depends primarily on the size of the request and its result rather than on the volume of underlying data, details in Section~\ref{sec:coordinate-to-reduce-network}. 

Consider an automotive manufacturer operating robotic fastening cells across multiple production facilities. Each facility stores its fastening events locally using the logical table \texttt{automotive\_fastening\_events}. An application can submit the following query to any authorized AnyLog agent:

\begin{Verbatim}[commandchars=\\\{\}]
\textbf{SELECT} 
plant_id, COUNT(*) \textbf{AS} failed_fastenings, AVG(torque_nm) \textbf{AS} average_failed_torque
\textbf{FROM} automotive_fastening_events
\textbf{WHERE} event_timestamp >= TIMESTAMP '2026-07-12 15:30:00' AND result_code = 'FAIL'
\textbf{GROUP BY} plant_id;
\end{Verbatim}

Using the Distributed Metadata Layer (Section~\ref{sec:metadata}) AnyLog resolves the \emph{table policy} for \texttt{automotive\_fastening\_events}. This policy defines the logical table and identifies the authorized AnyLog agents that host relevant data. Similar to MapReduce~\cite{dean2008mapreduce}, AnyLog then executes the query concurrently across those agents. Each participant filters its local records and computes a partial result containing the failure count, torque sum, and number of torque observations. These compact intermediate results are returned and combined to produce the global failure count and average failed torque for each plant.

The query operates across the manufacturer without first transferring every fastening event to a central historian, enterprise database, or cloud data lake. Each cell can contribute observations as soon as they are captured, so query freshness depends primarily on local processing latency rather than the completeness of centralized ingestion pipelines, and only the request and partial results traverse the network. As data volumes grow, ingestion, storage, and computation scale horizontally across additional AnyLog agents. Separate agents may collect different streams and perform filtering, aggregation, event detection, preprocessing, or inference in parallel, while computationally intensive workloads are assigned to agents with greater processing capacity.

The coordination benefit is as important as the computational benefit. A manufacturer operating six plants with eight fastening cells per plant would typically need to maintain 48 endpoint definitions, credential bindings, schema mappings, broker topics, and failover procedures. With AnyLog, the application integrates once with the logical table \nolinkurl{automotive\_fastening\_events} or a Unified Namespace pattern such as \nolinkurl{Plant / Area / Line / Cell / *}. When a new cell is added, it publishes its metadata and policies and becomes discoverable through the existing interface, eliminating the need for a new application-level integration.

\subsection{Coordinate the Fabric Through Shared Metadata}
\label{sec:coordinate-to-reduce-network}

A distributed platform requires a common understanding of the resources available across the system. Applications and AI agents must be able to discover datasets, determine which agents host relevant information, understand schemas, identify industrial assets and services, and evaluate security policies without requiring administrators to configure every participating agent manually. AnyLog addresses this requirement by separating the platform's shared knowledge from its operational data.

Operational data remains in the local databases managed by each AnyLog agent. Agents publish compact metadata that describes what data, services, resources, and policies they make available. This metadata forms a shared, dynamic operational knowledge layer through which users, applications, and AI agents can discover resources, coordinate and distribute work, enforce security and operational policies, and cooperate while allowing each participant to query the underlying data and metadata.

When an agent creates, adopts, updates, or removes a dataset, it publishes metadata describing the corresponding logical table, schema, physical location, ownership, Unified Namespace relationships, available services, and the policies governing access, security, retention, replication, and other lifecycle requirements, as discussed in Section~\ref{sec:metadata}. This metadata is dynamically discovered and synchronized across the Edge Data Fabric, allowing newly available or changed resources to become visible without requiring reconfiguration. AnyLog can therefore abstract the distributed platform behind a single logical API while reducing unnecessary broadcasts, centralized data movement, and overall network traffic.

For example, suppose that 1,000 industrial sensors each generate one 100-byte record per second. Excluding protocol overhead, continuously transmitting those records to a centralized repository would transfer approximately:

\[
1{,}000 \times 100 \times 86{,}400
= 8.64\ \text{GB per day}.
\]

Assume, conservatively, that each sensor is described by a 4-KB metadata record containing its identifier, schema, logical table, agent location, asset relationship, and access policy, and that every descriptor is retransmitted once per hour. Metadata synchronization would transfer approximately:

\[
1{,}000 \times 4{,}096 \times 24
= 98.3\ \text{MB per day}.
\]

Under these assumptions, this metadata-first coordination model reduces daily transmission by approximately (98.9%)~\cite{chang2008bigtable,leung2009spyglass}. In practice, metadata is generally transmitted only when definitions or operational conditions change, so the control-plane traffic may be substantially lower. The precise savings depend on the sampling rate, record size, update frequency, replication policy, and query-result volume.

The same effect appears at the workload level. Suppose 40 fastening cells each emit 100 records per second at 100 bytes per record, and a plant dashboard needs one-minute failure counts and average torque values. Shipping raw telemetry to a central service would move approximately

\[
40 \times 100 \times 60 \times 100
= 24\ \text{MB per minute}
\]

before protocol overhead. If the same request is executed locally, each cell can return only a few partial aggregates, reducing the response to the order of kilobytes. On fixed and local networks, such as Ethernet, fiber, and Wi-Fi, this reduces congestion; on metered cellular or satellite backhaul, it reduces recurring operating cost directly.

This approach follows established in-network aggregation techniques, which have shown that processing data near its source can reduce communication costs by an order of magnitude compared with centralized collection~\cite{madden2002tag}.

\subsection{Local Autonomy at Horizontal Scale Without Central Dependency}
\label{sec:horizontal-scale}

% AnyLog scales horizontally by distributing data, processing, and services across autonomous edge agents without making centralized infrastructure part of the critical path. Each agent manages its local data, executes workloads using its own resources, and coordinates with other agents through shared metadata. As new sites join, they contribute not only additional data but also compute, storage, and services, allowing platform capacity to expand with the deployment rather than concentrating demand at a central location.

% This model follows a principle established by peer-to-peer systems: discovery can be separated from where resources are stored and used. Napster used a shared index to locate files that remained on participating computers, while BitTorrent allowed peers to exchange file fragments directly and contribute bandwidth and storage to the network~\cite{ding2005peer,cohen2003incentives}. AnyLog applies this principle to operational computing by using shared metadata to discover distributed data, services, and computational resources while keeping storage and execution at their source. Through the Single System Image described in Section~\ref{sec:ssi}, applications and AI agents interact with these geographically distributed resources as one logical system~\cite{cattell2011scalable,shi2016edge}.

AnyLog scales horizontally by distributing data, processing, and services across independently operating edge agents. Each additional AnyLog agent  contributes its own data, compute, storage, and services, allowing platform capacity to grow without concentrating demand on centralized infrastructure.
Like peer-to-peer systems, it separates discovery from storage: Napster indexed files retained by participating computers, while BitTorrent allowed peers to exchange portions of a file concurrently and contribute partial network capacity~\cite{ding2005peer,cohen2003incentives}. 
AnyLog applies a related principle to operational computing by using shared metadata to discover distributed data, resources, and services while keeping storage and execution local to each agent.
As new AnyLog agents join, they contribute both workload and additional compute, storage, and services to the fabric. A Single System Image (Section~\ref{sec:ssi}) then presents these resources to applications and AI agents as one logical system~\cite{cattell2011scalable,shi2016edge}.

% AnyLog scales horizontally by distributing data, processing, and services across autonomously operating edge agents. Like peer-to-peer systems, it separates discovery from storage: Napster indexed files retained by participating computers, while BitTorrent allowed peers to exchange file pieces concurrently and contribute network capacity~\cite{ding2005peer,cohen2003incentives}. AnyLog applies a related principle to operational computing by using shared metadata to discover distributed data, resources, and services while keeping storage and execution at each location. As new sites join, they contribute both workload and additional compute, storage, and services to the fabric's single system image, as described in Section~\ref{sec:ssi}, allowing platform capacity support horizontal growth rather than accumulate centrally~\cite{cattell2011scalable,shi2016edge}.

New sites and devices may publish data to an existing agent or deploy additional AnyLog agents. Once a new agent publishes its metadata and policies, its data, services, and resources become discoverable through the fabric's existing interfaces without requiring applications to be reconfigured. Each added agent also contributes compute, memory, storage, and services, allowing platform capacity to grow horizontally alongside the workload rather than accumulating on centralized infrastructure:
\[ \mathbf{C}_{\mathrm{platform}} = \sum_{i=1}^{n}\mathbf{C}_i, \] where \[ \mathbf{C}_i = \left( C_i^{\mathrm{compute}}, C_i^{\mathrm{memory}}, C_i^{\mathrm{storage}}, C_i^{\mathrm{services}} \right) \]

Capacity therefore expands across operational sites rather than accumulating on a central gateway, server, or cluster~\cite{cattell2011scalable,shi2016edge}. Unlike centralized scale-out platforms that must absorb data from every remote site~\cite{dewitt1992parallel}, AnyLog makes each site part of the computing platform. Agents ingest and process data locally, execute requests concurrently, and return partial results that a Query Coordinator combines into a unified response.

Cooperation does not create a dependency on continuous global connectivity. Each agent continues collecting data, executing local queries, enforcing security policies, running automation, and supporting local applications without a central coordinator or cloud service. During a disruption, the fabric contracts to the agents reachable within the current network partition; connected agents continue coordinating, while unreachable agents become temporarily excluded. This allows queries to remain \emph{partially} satisfied by connected replicas, avoiding the complete operational disruption common with centralized architectures~\cite{shi2016edge,iorga2018fog}. Associated with each query result is a \emph{query summary} that includes which agents participated and what data may be missing. Once connectivity returns, agents refresh their metadata and resume broader coordination without restarting local workloads. This is particularly valuable in denied, disconnected, intermittent, and low-bandwidth environments~\cite{dod2020c3modernization}.

\emph{Retention policies} complete the model. Current observations remain near the physical process while they are most valuable, while older records, summaries, and selected events may be offloaded or replicated to higher-capacity agents, object stores, historians, or cloud platforms~\cite{ali2021hidden}. Edge storage and computation remain focused on current operations and local AI inference, while cloud resources continue supporting training, long-term retention, reporting, and enterprise analytics~\cite{singh2023edge,meuser2024revisiting,surianarayanan2023survey,gill2025edge,kusiak2018smart,wan2020artificial}.

\subsection{Integrate Through Open Standards and MOSA Principles}

AnyLog is designed in alignment with the Department of Defense's Modular Open Systems Approach (MOSA), which promotes modular architectures, open interfaces, interoperability, and the ability to integrate or replace components without redesigning the complete system~\cite{dodMOSA2025}. AnyLog applies these principles through a common agent runtime, configurable node roles, standards-based interfaces, and loosely coupled services that can be deployed, upgraded, or expanded independently.

Through southbound interfaces such as OPC UA~\cite{mahnke2009opc}, MQTT~\cite{mishra2020use}, Modbus~\cite{thomas2008introduction}, REST~\cite{murali2019hands}, gRPC~\cite{wang1993grpc,grpcDocumentation}, and standard database connectors, AnyLog captures data from industrial equipment, control systems, historians, databases, and external applications. Through northbound interfaces such as SQL, REST APIs, gRPC, and the Model Context Protocol (MCP)~\cite{hou2025model}, it exposes distributed data and services to applications, analytics platforms, automation systems, and AI agents.

This modular, standards-based architecture allows AnyLog to operate as an independent data layer across heterogeneous hardware, operating systems, networks, and deployment environments. Organizations can introduce AnyLog incrementally, retain existing systems of record and operational workflows, and integrate new technologies through open interfaces rather than proprietary point-to-point dependencies.

% \section{Data and Context Integration and Unified APIs}
\section{Bridging the Physical Edge to Applications and AI}
\label{ch:integration}

% \subsection{Connecting Operational Systems}

% The value of an industrial data platform depends on its ability to integrate with existing operational systems. Manufacturing equipment, sensors, programmable logic controllers (PLCs), historians, enterprise databases, cloud platforms, and business applications all contribute to an organization's operational picture. These systems have evolved over many years and typically use different communication protocols, data models, and storage technologies.

% Replacing this infrastructure is rarely practical or desirable. Instead, the \textbf{AnyLog Edge Data Fabric} operates as an independent distributed data layer above existing systems. It interfaces directly with historians, industrial control systems, databases, cloud platforms, and enterprise applications while presenting applications and AI with a single logical interface to operational data distributed across edge locations, historians, and local databases.
% Integration therefore becomes more than connecting devices and applications. It transforms independent operational systems into participants of a unified distributed platform.

\subsection{Automatic Schema Creation and Management}
\label{sec:auto-schema}

One of the most persistent challenges in distributed industrial environments is schema fragmentation. Equipment, applications, and databases are often configured independently by different vendors, integrators, and facilities, causing the same operational data to be represented through different table structures, field names, units, and data types. Preventing these inconsistencies typically requires extensive coordination across teams and sites, slowing deployment, integration, and future expansion.

AnyLog addresses this through automatic schema creation and distributed management. When new data is ingested, the receiving agent consults the Distributed Metadata Layer to determine whether an existing \emph{table policy} describes it. If a matching policy exists, the agent adopts the established schema. Otherwise, it generates a schema from the incoming data and publishes the new definition to the metadata layer, making it immediately discoverable and reusable by other agents receiving similar data.

This process allows new data sources to join the fabric through a plug-and-play model without requiring a separate integration project. Ingested data, physical observations, and AI-generated predictions, classifications, recommendations, and other insights remain stored locally, while their schemas, locations, relationships, and access policies become discoverable through the metadata layer. Existing definitions can then be reused as the deployment expands, maintaining a consistent platform-wide data model regardless of scale.

% Formally, a schema can be modeled as
% \[
% s = (A_s,\tau_s,\kappa_s)
% \]
% where \(A_s\) is the attribute set, \(\tau_s : A_s \rightarrow \mathcal{T}\) assigns a data type to each attribute, and \(\kappa_s\) captures constraints such as nullability, units, accepted ranges, indexing rules, and other validation requirements. A record \(r\) conforms to schema \(s\), written \(r \models s\), when the required attributes are present and the values satisfy the declared types and constraints.

% Let \(\operatorname{sig}(r)\) denote the structural signature inferred from an incoming record or batch. An existing schema \(s\) is reusable for \(\operatorname{sig}(r)\) when the incoming signature is compatible with \(s\), written \(\operatorname{sig}(r) \preceq s\). In practical terms, compatibility means that required attributes are preserved, shared attributes remain type-compatible, and declared constraints are not violated. Exact matches and backward-compatible extensions may therefore reuse the same logical schema, while incompatible type changes or missing required attributes cause the agent to publish a new schema identifier and, if necessary, a new logical table version.

For example, an AnyLog agent monitoring a robotic fastening cell in an automotive assembly plant may register the logical table name \texttt{automotive\_fastening\_events} together with the following schema in the Distributed Metadata Layer:

\begin{verbatim}
table_name: automotive_fastening_events

cell_id: str
station_id: str
robot_id: str
vehicle_model: str
operation: str
event_timestamp: datetime
torque_nm: float
angle_deg: float
cycle_time_ms: int
result_code: str
fault_code: str | null
\end{verbatim}

The metadata definition may also specify units of measurement, required fields, accepted value ranges, indexing rules, retention policies, and relationships among the production cell, station, robot, vehicle model, and assembly operation.

A different agent may later receive the following JSON object from a newly commissioned fastening cell:

\begin{verbatim}
{
  "cell_id": "CELL-27",
  "station_id": "ST-04",
  "robot_id": "RB-112",
  "vehicle_model": "EV-X",
  "operation": "battery_tray_fastening",
  "event_timestamp": "2026-07-12T15:42:18.421Z",
  "torque_nm": 86.4,
  "angle_deg": 137.2,
  "cycle_time_ms": 1840,
  "result_code": "PASS",
  "fault_code": null
}
\end{verbatim}

Before storing the object, the receiving agent derives its structural signature and consults the Distributed Metadata Layer for a compatible schema. It compares the incoming field names and value types against the registered definition, confirming, for example, that \texttt{cell\_id} is a string, \texttt{event\_timestamp} is a datetime, \texttt{torque\_nm} is a floating-point value, and \texttt{cycle\_time\_ms} is an integer.

When the schema matches, the agent adopts the existing logical table, creates the corresponding local storage structure, validates or converts the incoming values, and declares through the metadata layer that it contains compatible data.

An application can then issue a query such as:

\begin{Verbatim}[commandchars=\\\{\}]
\textbf{SELECT} cell_id, station_id, robot_id, result_code,
\textbf{FROM} automotive_fastening_events
\textbf{WHERE} event_timestamp >= NOW() - 5 minutes \textbf{AND} result_code = 'FAIL';
\end{Verbatim}
which is resolved through the logical table name and distributed to every authorized agent that has declared data conforming to the \texttt{automotive\_fastening\_events} schema. This includes the agent that originally registered the schema and any newly commissioned agents that adopted it. Each agent evaluates the query against its local data, returns its partial result set, and the Distributed Query Engine combines the partial results into one unified response.

The new production cell becomes immediately available to existing distributed SQL queries, dashboards, quality-control applications, and AI models without manual schema replication, centralized data movement, or changes to application logic. As deployments grow to thousands of edge agents, new data sources automatically join the fabric, allowing applications and AI models to continue operating without reconfiguration.

% Platforms such as Splunk reduce some of this fragmentation through common information models and source-specific add-ons~\cite{splunkCIM,splunkAddons}. AnyLog pushes the unification further: schemas, logical tables, data locations, agent capabilities, asset relationships, and governing policies are represented as distributed metadata and made available through the same architecture as the records themselves. Applications therefore integrate once with the Edge Data Fabric and use metadata-driven discovery in place of hard-coded database locations, source-specific routing logic, and separate discovery mechanisms for every operational system.

\subsection{Dynamic Operational Context Through Unified Namespaces}

Relational tables organize operational data into rows and columns that applications can query using SQL, but they do not fully represent the physical systems, assets, and relationships that give those records meaning.
AnyLog complements relational data models with one or more \emph{logical} \textbf{Unified Namespaces (UNS)} that organize distributed assets according to the operational environment and how users need to access the data, independent of how the data was captured or where it is physically stored.

A UNS represents assets through a root, parent, and child hierarchy. In a manufacturing environment, this structure may follow:

\begin{quote}
\texttt{Plant / Area / Line / Cell / Station / Asset / MeasurementOrService}
\end{quote}

\emph{UNS policies} published through the Distributed Metadata Layer define how each semantic path maps to the relevant tables, columns, filters, services, agents, databases, and connectors. This allows operators, dashboards, MES applications, quality systems, and AI agents to navigate the same operational environment according to their respective workflows and analytical needs. One user may organize data by plant, line, cell, and station, while another maps the same records by robot type, product family, maintenance condition, quality state, or utility resource. Each view resolves to the same distributed data without requiring the underlying storage or management to be changed or duplicated.

For example, the path

\begin{quote}
\texttt{CarPlant / BodyShop / Line3 / Cell27 /}\
\texttt{FasteningStation04 / Torque / CurrentValue}
\end{quote}

\begin{verbatim}
{
  "uns": {
    "name": "FasteningStation04/Torque/CurrentValue",
    "namespace":"MI/BodyShop/Line3/Cell27/FasteningStation04/Torque/CurrentValue",
    "uns_level": "measurement",
    "dbms": "automotive",
    "table": "automotive_fastening_events",
    "column": "torque_nm",
    "data_type:unit": "float:Nm",    
    "parent": "fastening_station_04_policy_id"
  }
}
\end{verbatim}

The policy links the semantic path to the \nolinkurl{torque\_nm} column in the logical table \nolinkurl{automotive\_fastening\_events} and applies the filters identifying the corresponding plant, area, line, cell, and station. It also supplies contextual information such as the data type, engineering unit, measurement type, and parent relationship. Applications can therefore request the current torque value through the UNS path while AnyLog resolves the underlying data source and query automatically.

The same distributed records can support multiple Unified Namespaces and levels of analysis, from a single asset or production line to conditions spanning multiple facilities. Production teams may organize data by plant, line, and cell; maintenance systems by equipment type or condition; and quality applications by product or process stage. 
Unlike MQTT-centric namespace designs, where operational views are coupled to broker topics and topic hierarchies~\cite{standard2019mqtt,mishra2020use,eclipseSparkplug2022}, AnyLog defines semantic views through metadata policies. This allows multiple namespaces to organize the same underlying data without republishing, duplicating, or relocating records. Because these semantic paths remain stable as agents, databases, sites, communication protocols, and broker topologies evolve, applications and AI agents can continue navigating and querying the data through their established namespace paths without modification.

\subsection{Projecting UNS Structures as a Knowledge Graph}

\begin{figure}[t]
\centering
\includegraphics[width=.8\linewidth]{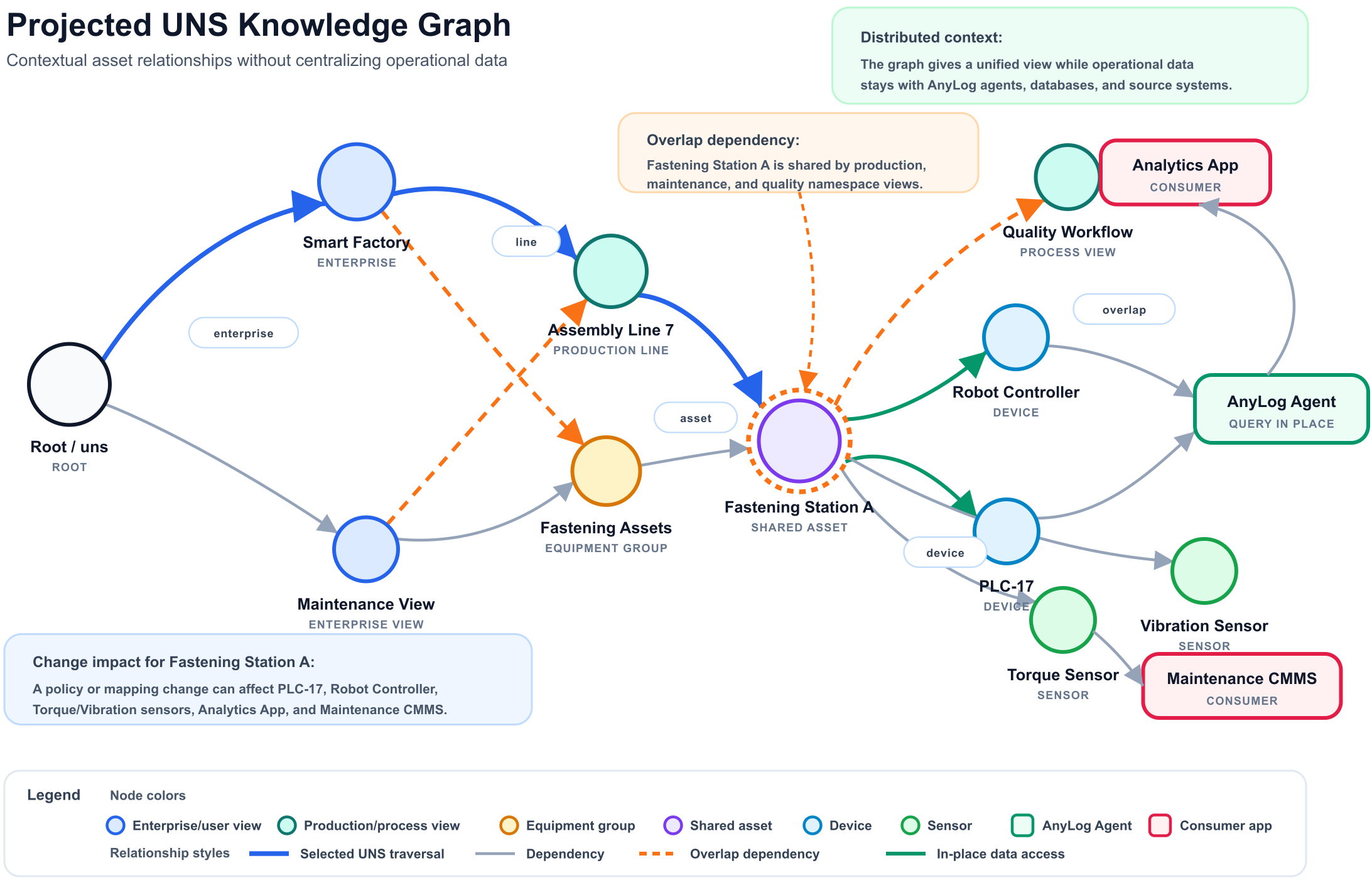}
\caption{Projected UNS knowledge graph for an industrial fastening station. The orange dashed ring and paths identify an overlap dependency: Fastening Station A is referenced across production, maintenance, and quality namespace views. By making this shared context explicit, the graph helps users understand where an asset is used, which systems depend on it, and what may be affected before a policy, mapping, equipment, or data-model change is made. In this example, a change to Fastening Station A can be traced across upstream namespace context and downstream dependencies, including PLC-17, Robot Controller, Torque/Vibration sensors, Analytics App, and Maintenance CMMS. This provides enterprise-wide impact visibility without centralizing operational data, since AnyLog brings the query to the data and only the result sets back.}
\label{fig:uns-knowledge-graph}
\end{figure}

A Unified Namespace organizes assets along hierarchical operational paths, but hierarchy alone cannot fully represent many-to-many relationships, cross-functional dependencies, or assets that participate in multiple contexts. Projecting UNS policies as a \textbf{Knowledge Graph} extends these paths into an interconnected model of the operational environment, making relationships among assets, measurements, services, applications, users, and policies directly visible~\cite{ji2021survey}.

Figure~\ref{fig:uns-knowledge-graph}, for example, shows a fastening station shared across production, maintenance, and quality views. Connecting these perspectives through the same asset allows users and AI agents to move across operational contexts, discover related data and services, and determine how changes may affect sensors, controllers, applications, and workflows, which improves discovery, dependency analysis, and safer change management.

The strategic importance of industrial knowledge graphs is reflected in Schneider Electric's 2026 agreement to acquire Cognite for $3.1$~billion~\cite{schneider2026cognite}. Cognite Data Fusion creates this context by ingesting and transforming industrial data within a cloud-based platform. AnyLog delivers a similar view but notably through a distributed architecture: UNS policies maintained in the Distributed Metadata Layer define the knowledge graph, while operational data remains managed and hosted at the edge by AnyLog agents.

Because relationships are defined \emph{logically} through metadata policies, AnyLog can expand and update the graph without first consolidating the underlying data. Enterprises gain a shared contextual model while avoiding unnecessary duplication, synchronization delays, network demands, and centralized infrastructure costs. As new assets, applications, users, and relationships publish UNS policies, the graph expands through a plug-and-play process while the operational data remains distributed at the edge.

\subsection{Southbound Integration}
\label{sec:southbound}

Operational data enters the AnyLog platform through \textbf{Southbound Connector Services}. These services establish communication with industrial devices, applications, and databases using industry-standard protocols and interfaces.

The connector framework is intentionally modular. Each AnyLog agent enables only the connectors required for its deployment, allowing the same software platform to operate efficiently on embedded gateways, industrial servers, enterprise systems, or cloud infrastructure. Connectors are activated through configuration, enabling organizations to adapt deployments without modifying application code.

The platform includes native support for widely adopted industrial and enterprise technologies such as OPC UA, MQTT, Modbus, REST, gRPC, relational databases, object storage, file-based interfaces, and RTSP for live video streaming. Because the framework is open, organizations can also integrate proprietary systems or leverage third-party connectors when required.

Regardless of the protocol, every connector performs the same architectural function. It acquires operational information and streams it to the local AnyLog agent. From that point forward, the ingestion process is fully automated and governed by policies. Incoming data is normalized, existing schemas are reused or new schemas are generated automatically when not provided. Policies determine how the data is retained, archived, replicated, aggregated, monitored, and secured, while making it available through the platform's unified APIs. Once ingested, the original communication protocol becomes largely irrelevant---the data is now part of the distributed platform and can be consumed consistently by applications, administrators, and AI regardless of how it was acquired.

\subsection{Northbound Interfaces For Enterprise Integration}
\label{sec:northbound}

Applications, dashboards, enterprise systems, analytics platforms, automation tools, and AI agents interact with the Edge Data Fabric through a common set of \textbf{northbound interfaces}. SQL and REST provide familiar query and application access, gRPC and Kafka~\cite{kafka} support event-driven integration, and MCP allows AI agents to discover and invoke distributed data, tools, services, and contextual resources. Through these same interfaces, users, applications, and AI agents can query data stored across the edge and in connected historians through one API. Within the fabric, a historian is represented as \emph{another edge} through additional \emph{table policies}, described in Section~\ref{sec:metadata}, allowing current and archived data to be accessed through a single edge data platform.

Since distributed data and metadata are exposed through a single API, SQL queries and native AnyLog commands can be issued using standard tools such as Postman~\cite{kore2022api} or integrated directly into applications. AnyLog also includes the \textbf{AnyLog Edge Data Manager (EDM)}~\cite{anylogEDM}, which provides administrators and developers with a graphical Single System Image for exploring, querying, testing, monitoring, configuring, and operating the distributed platform.
Through the same GUI interface, users can navigate asset relationships in the Unified Namespace, compare metrics stored across multiple agents, interact with the fabric through the native MCP client, and inspect returned records and files.

\section{Operating the Distributed Platform}
\label{chap:operating-platform}

\subsection{Simplifying Distributed Operations through a Single System Image}
\label{sec:ssi}

Deploying a distributed edge system is only the beginning of the operational challenge. As hundreds or thousands of gateways, servers, and industrial computers come online, administrators must determine where data resides, monitor platform health, diagnose failures, distribute policies, enforce security, and coordinate services while tracking device-specific addresses, configurations, software versions, databases, credentials, and dependencies. In conventional deployments, these point-to-point responsibilities multiply with every new system, causing management complexity to grow faster than the engineers available to manage it. Rather than operating a collection of silos, AnyLog provides a \textbf{Single System Image} through which administrators, applications, and AI agents interact with distributed data, compute resources, and services as though they belonged to one unified, cloud-like system.

Through a unified interface, administrators can:

\begin{itemize}
\item discover agents, roles, datasets, metadata, and services;
\item inspect databases, schemas, policies, and data locations;
\item monitor agent health and resource utilization;
\item track replication, aggregation, and automation activity; and
\item execute distributed queries and perform administrative operations.
\end{itemize}

For example, an automotive manufacturer may operate 600 AnyLog agents across assembly plants, supplier facilities, test environments, and distribution centers. The SSI presents these resources as one operational platform while allowing administrators to filter the environment by facility, production line, agent role, software version, available service, or current state.

The SSI can also provide the operational foundation for an edge \textbf{Digital Twin}~\cite{liu2021review,fuller2020digital,singh2021digital,tao2022digital}. Because agents publish metadata describing their identities, capabilities, datasets, services, configurations, policies, and availability, the SSI can reproduce the logical structure and observable state of the distributed environment without centralizing its underlying operational data.

The metadata state can be copied into a development environment, digital twin, or newly deployed system using:

\begin{verbatim}
blockchain seed from [ip:port]
\end{verbatim}

This command initializes an authorized AnyLog agent outside the production environment with the schemas, policies, asset relationships, services, and resource locations known to the production fabric. Users can therefore reproduce the logical structure of an existing deployment, test configurations and workflows locally, and later apply the same metadata-defined roles and deployment instructions to production agents. When permitted by policy, the replicated environment may also receive a limited subset of operational data---for example, only 100 rows or records from an approved historical time range---allowing a functional digital twin to be created quickly without exposing the complete production dataset. The SSI remains the common interface for observing and managing both the modeled environment and the deployed physical edge.

% \subsection{Unified Management Independent to Deployment Size}

% Administrative operations use the same logical model regardless of deployment size. An operation may target a single agent, a group selected through metadata, or every authorized participant in the platform.

% For example, an administrator may distribute a new retention policy to all agents assigned to automotive paint lines, enable an aggregation service only on agents that collect energy data, or request diagnostic information from every backup agent in a particular geographic region. The administrator selects the intended capabilities or operational scope, and the platform resolves the eligible agents.

% Although every agent executes the same AnyLog software platform, agents may be configured for different responsibilities. Some collect and store operational data, some expose northbound APIs and coordinate distributed queries, some provide backup capacity, and others monitor the health and utilization of the wider deployment. Management therefore focuses on roles and capabilities across one shared runtime.

\subsection{Configuration-Driven, One-Click Deployment}
\label{sec:config}

Industrial integration is traditionally implemented as a series of custom software projects: building connectors, mapping schemas, exposing APIs, and repeating the process for every new site, device, or application. As deployments expand, these one-off integrations accumulate into complex, brittle systems that are costly to maintain.

AnyLog replaces custom integration with a configuration-driven deployment model. Reusable configuration files and metadata \emph{Node Policies} (Section~\ref{sec:metadata}) define each agent's roles, connectors, data sources, schemas, interfaces, services, network settings, and security policies. The same runtime can therefore be deployed at the edge, in operations centers, or in cloud environments, with each instance configured from a shared knowledge base rather than implemented as a separate integration project.

Once defined, configurations can be deployed and replicated automatically across geographically distributed sites~\cite{anylogDeploymentScripts}. The same model integrates with orchestration platforms such as IBM Open Horizon~\cite{lfedge_open_horizon}, IBM Edge Application Manager (IEAM)~\cite{ibmIEAMOverview}, and Dell Distributed Private Cloud (formerly NativeEdge)~\cite{dell_nativeedge}. Configuration and policy updates are published once through the Distributed Metadata Layer and automatically discovered by the relevant agents, while local policies allow individual sites to accommodate differences in equipment, regulations, network conditions, and security requirements.

\subsection{Observability and Diagnostics}

Operating a distributed data platform requires visibility into both the fabric as a whole and the individual systems that compose it. AnyLog observability spans agents, databases, datasets, services, policies, and distributed operations. Administrators can locate logical tables, inspect metadata declarations, monitor replication and backup activity, review aggregation and automation processes, and trace the execution of distributed queries across participating agents.

Using the \texttt{query status} command, users can trace a distributed query and identify:
\begin{itemize}
    \item the agent that coordinated the request;
    \item the metadata used to select participants;
    \item the agents to which the request was distributed;
    \item the execution time reported by each participant;
    \item the number of records or partial aggregates returned;
    \item agents that were unavailable or rejected the request; and
    \item the operations used to combine the partial results.
\end{itemize}

For example, a query may execute against production data stored at 27 facilities. If 26 agents respond within milliseconds but one requires several seconds, the \texttt{query status} command allows the user to identify the slow agent. Applications may also configure the \texttt{run client} SQL command with \texttt{subset=true} and \texttt{timeout=2 seconds}, instructing AnyLog to return a unified view of the available partial results when an agent does not respond within two seconds, rather than delaying the entire query:

\begin{verbatim}
run client () sql [dbms] subset=true and timeout=2 seconds
SELECT ...
\end{verbatim}

The returned response identifies the agents that participated and those that were unavailable or timed out, allowing the application to determine whether the result is complete. Additional diagnostic services allow administrators to verify connectivity, inspect local resource utilization, analyze metadata synchronization, review policy execution, and examine platform activity without establishing a separate administrative session with every machine.

\subsection{Policy-Governed Multi-Organization Collaboration}

Most deployments use AnyLog to unify resources within one organization. The same architecture can also support controlled collaboration among manufacturers, suppliers, equipment vendors, logistics providers, utilities, contractors, and customers.

Collaboration is optional and policy-driven. Each organization operates its own agents and selectively publishes metadata describing the datasets, services, or derived information it chooses to make available. Other participants can discover those resources, but access remains subject to the policies enforced by the organization that owns them.

For example, an automotive manufacturer may allow a robotic-equipment supplier to access failure summaries for the supplier's installed robots. The supplier can analyze the authorized information across multiple plants without receiving the manufacturer's complete production database or gaining administrative control over plant infrastructure.

Policies can also define the permitted depth of collaboration. One partner may be allowed only aggregate SQL such as \texttt{COUNT}, \texttt{AVG}, \texttt{MAX}, or anomaly summaries across all cells using its equipment. Another may be allowed row-level \texttt{SELECT} statements on approved logical tables but only for whitelisted columns such as \texttt{timestamp}, \texttt{fault\_code}, \texttt{torque\_nm}, and \texttt{robot\_id}. Namespace access can be constrained in the same way: a supplier may be allowed to navigate a UNS branch such as \texttt{CarPlant / BodyShop / Line3 / RobotVendorX / *} while being unable to inspect unrelated lines, product models, utilities, or columns outside that collaboration boundary.

Organizations determine:

\begin{itemize}
    \item which resources are discoverable;
    \item which identities may request them;
    \item which query classes are permitted, such as approved aggregations or row-level \texttt{SELECT} statements;
    \item whether access applies to raw records or derived results;
    \item which UNS branches, logical tables, or columns are visible to a partner;
    \item which operations may be executed;
    \item how frequently information may be requested; and
    \item whether results may leave the local environment.
\end{itemize}

Multiple organizations can therefore deploy AnyLog independently and still cooperate through a shared logical platform while preserving their own databases and ownership boundaries. The data owner continues to enforce the policy boundary at the source agent, while the consumer experiences only the approved logical view.

\subsection{Distributed Security and Local Policy Enforcement}

AnyLog does not treat network membership as proof of trust. Each agent is assigned a cryptographic identity managed in software or protected by a hardware-based \textbf{Trusted Platform Module (TPM)}~\cite{tomlinson2017introduction}. Agents use these identities to authenticate requesters and participating systems, establish encrypted channels, protect data in transit, and enforce access policies where each protected resource resides~\cite{rose2020zero}.

\textbf{Access Policies} define which identities may access specific services, databases, tables, and time ranges. Published through the Distributed Metadata Layer, they provide a shared view of authorized access while preserving local enforcement. Before executing a service or returning data, the agent responsible for the resource authenticates the requester and verifies that an applicable policy permits the request. A Query Coordinator may discover resources and coordinate execution, but it cannot expand or override the permissions enforced by the agents hosting them.

This design also limits the impact of a compromised identity. Stolen credentials may expose resources already authorized for that identity~\cite{mitre2025validaccounts}, but they do not grant fabric-wide access or bypass policies enforced by other agents. The potential damage remains bounded by the compromised identity's assigned permissions, and revoking those permissions prevents subsequent access~\cite{rose2020zero,mitre2025validaccounts}.

\subsubsection{Security Policy Model}
\label{sec:policy-model}

AnyLog uses a whitelist-based authorization model. Access policies published through the Distributed Metadata Layer define which authenticated users, applications, and AI agents may access specific services and data resources; requests without a matching policy are denied. For example, an equipment supplier may query fault counts for the machines it maintains but not access raw production data or unrelated equipment. The agent hosting the data authenticates the requester and enforces the permitted scope, ensuring that authorization is controlled by the resource owner rather than the requester.

Let a request be represented as

\[
x = (i,s,d,\mathcal{T},\Delta),
\]

where \(i\) is the authenticated requester identity, \(s\) is the requested
AnyLog service, \(d\) is the requested database, \(\mathcal{T}\) is the set of
requested tables, and \(\Delta\) is the requested time range. A component may
be omitted when it is not applicable to the requested service.

An access policy published through the Distributed Metadata Layer is represented
as

\[
p = (S_p,G_p,\Delta_p),
\]

where \(S_p\) is the set of authorized requester identities, \(\Delta_p\) is
the permitted time range, and

\[
G_p
\subseteq
\mathcal{S}
\times
\mathcal{D}
\times
2^{\mathcal{T}}
\]

is the set of authorized service $\mathcal{S}$, database $\mathcal{D}$, and table scopes $2^\mathcal{T}$. Each element

\[
(s_p,d_p,\mathcal{T}_p) \in G_p
\]

defines one permitted combination. The table scope \(\mathcal{T}_p\) may
identify one table, a subset of tables, or all tables within the authorized
database.

A policy \(p\) authorizes request \(x\) when

\[
i \in S_p,
\]

and there exists an authorized scope

\[
(s_p,d_p,\mathcal{T}_p) \in G_p
\]

such that

\[
s=s_p,
\qquad
d=d_p,
\qquad
\mathcal{T}\subseteq\mathcal{T}_p,
\qquad
\Delta\subseteq\Delta_p.
\]

Let \(\mathcal{P}_{M}\) denote the access policies available through the
Distributed Metadata Layer. The authorization decision is therefore

\[
\operatorname{Allow}(x)
\iff
\exists p\in\mathcal{P}_{M}
\text{ such that }p\text{ authorizes }x.
\]

If no policy authorizes the complete request, the request is denied:

\[
\neg\operatorname{Allow}(x)
\iff
\nexists p\in\mathcal{P}_{M}
\text{ such that }p\text{ authorizes }x.
\]

For distributed queries, the Query Coordinator identifies the agents holding
the requested data and forwards the authenticated request context to them.
Each participating agent evaluates the applicable policies obtained through the
Distributed Metadata Layer before executing the service or satisfying the request. The
Query Coordinator thus coordinates execution and combines authorized results, but
it cannot grant access or override the authorization decision enforced where
the requested resource resides.

This initial model focuses on authorization for services, databases, tables,
and time ranges. The same whitelist-based structure can be extended to govern
columns, row predicates, Unified Namespace branches, permitted aggregation
functions, result formats, model artifacts, file services, and whether data may
be returned as raw records or only as policy-approved summaries.

\section{Building an AI-Native Edge Data Platform}
\label{chap:ai-native-platform}

\subsection{The Operational Foundation for Industrial AI}

The primary barrier to industrial AI is no longer access to capable models; it is giving those models immediate, governed access to operational data, metadata, and services. To monitor equipment, diagnose failures, coordinate production, and initiate actions safely, AI must understand current conditions, schemas and units, asset relationships, available services, system state, and the policies governing what it may access and execute. Today, this context remains fragmented across historians, databases, cloud platforms, asset-management systems, and application-specific APIs, forcing each new AI application to depend on custom connectors, synchronized datasets, and static schema mappings.

One consequence is \emph{agent sprawl}~\cite{gartner2026agentsprawl,elsayed2026agentic}. Without a unified interface to the operational environment, developers deploy specialized agents for individual systems and tasks, each with its own connectors, credentials, schema mappings, and application logic. One agent may query historian data, another access MES or ERP systems, another retrieve maintenance records, and others invoke control services or enterprise applications.

As operational systems expand, the number of agents, connectors, and synchronization workflows grows with them, increasing deployment complexity, management overhead, and security exposure. Each agent also retains only a partial view of the environment, making coordinated reasoning, consistent governance, and uniform access control increasingly difficult.

A scalable approach to industrial AI requires a shared operational foundation that can expand across applications, agents, assets, and sites. AnyLog provides this foundation through the \textbf{Edge Data Fabric}, which presents distributed data, metadata, compute resources, services, and policies as one governed logical environment. The \textbf{Distributed Metadata Layer} identifies available resources, their locations, relationships, and access rules; the \textbf{Virtual Data Lake} provides a common data model; the \textbf{Unified Namespace} adds operational context; the \textbf{Distributed Query Engine} executes computation where the data resides; and the \textbf{Single System Image} makes the expanding environment manageable as one platform.

With this foundation, new applications and AI agents can build on the same data, metadata, services, and policies instead of introducing another isolated integration stack. As assets, sites, datasets, services, and models join the fabric, they become available through the same logical interfaces, allowing industrial AI to expand across the enterprise without needing specialized middleware connectors, duplicated mappings, and synchronization workflows~\cite{kusiak2018smart,wan2020artificial}.

\subsection{Robust Decision-Making for Nonstationary Physical Systems }

As AI assumes greater responsibility for physical operations, the fault and attack surface changes. Decisions may depend not only on human-in-the-loop inputs, which can be delayed, incorrect, or manipulated, but also on observations generated by autonomous robots, vehicles, machines, and sensors that may be noisy, faulty, compromised, or operating from different perspectives. Because these inputs affect movement, coordination, production, and safety, they must be evaluated at the edge and protected against individual errors before they influence the critical decision path~\cite{zhou2019edge}.

Low-latency access alone does not ensure that a decision accurately reflects the physical environment. Multiple robots may observe the same event differently because of sensor noise, changing positions, communication delays, equipment faults, human error, or malicious behavior. Where redundant observations exist, fault-tolerant decision-making must combine them without allowing one incorrect input to lead to a consequential outcome.

Byzantine agreement enables consistent decisions despite faulty or malicious inputs~\cite{dolev1986reaching,lamport1983weak}, but traditional protocols often depend on repeated coordination and eventual convergence. Emerging approaches such as Proximal Byzantine Agreement (PBA) make robust edge decision-making more practical through one-shot statistical inference that bounds faulty influence while also quantifying decision uncertainty~\cite{shadmon2025enhancing,shadmon2025brief}.

AnyLog provides the distributed data-access layer needed to apply these methods with the Unified Namespace paths:
\nolinkurl{CarPlant/BodyShop/Floor/Robot-[A|B|C]/Objects/Forklift-17/Position}
The associated UNS policies resolve these paths to the relevant logical table, fields, and filters. A decision process can retrieve the latest observations through a distributed query:

\begin{Verbatim}[commandchars=\\\{\}]
\textbf{SELECT}
robot_id, object_id, position_x, position_y, confidence, observation_timestamp
\textbf{FROM} robot_object_observations
\textbf{WHERE} object_id = 'Forklift-17' 
\textbf{AND} observation_timestamp >= CURRENT_TIMESTAMP - INTERVAL '2' SECOND;
\end{Verbatim}

AnyLog identifies the agents holding the observations, retrieves the authorized records concurrently, and returns them as one logical dataset. PBA or another sensor-fusion protocol can then estimate the forklift's position while limiting the influence of noisy and faults~\cite{iyengar2004distributed,elmenreich2002introduction}.

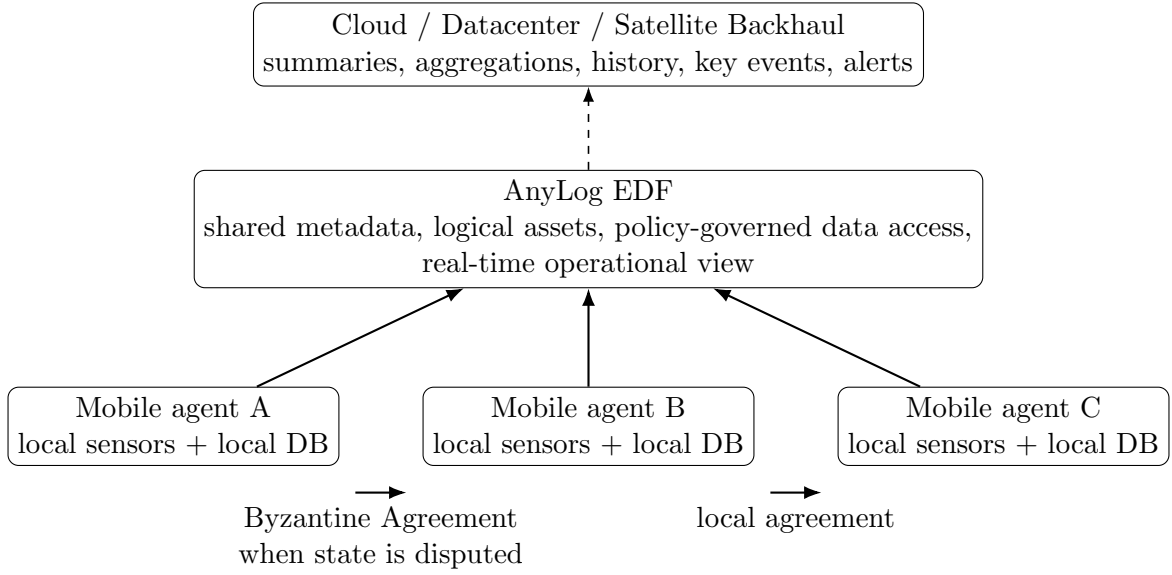
\begin{figure}
    \centering
    \begin{tikzpicture}[
        node distance=1.0cm and 1.1cm,
        box/.style={draw, rounded corners, align=center, minimum width=3.1cm, minimum height=1.15cm},
        smallbox/.style={draw, rounded corners, align=center, minimum width=2.6cm, minimum height=0.95cm},
        flow/.style={-Latex, thick},
        aux/.style={-Latex, dashed, semithick}
    ]
        \node[smallbox] (n1) {Mobile agent A\\local sensors + local DB};
        \node[smallbox, right=of n1] (n2) {Mobile agent B\\local sensors + local DB};
        \node[smallbox, right=of n2] (n3) {Mobile agent C\\local sensors + local DB};

        \node[box, above=1.3cm of n2, minimum width=7.2cm, minimum height=1.5cm] (edf) {AnyLog EDF\\shared metadata, logical assets, policy-governed data access, \\real-time operational view};
        \node[smallbox, above=1.1cm of edf] (cloud) {Cloud / Datacenter / Satellite Backhaul\\summaries, aggregations, history, key events, alerts};

        \draw[flow] (n1) -- (edf);
        \draw[flow] (n2) -- (edf);
        \draw[flow] (n3) -- (edf);

        \draw[flow] ([xshift=6pt,yshift=-10pt]n1.south east) -- node[midway, below=4pt, fill=white, inner sep=1pt, align=center] {Byzantine Agreement\\when state is disputed} ([xshift=-6pt,yshift=-10pt]n2.south west);
        \draw[flow] ([xshift=6pt,yshift=-10pt]n2.south east) -- node[midway, below=4pt, fill=white, inner sep=1pt, align=center] {local agreement} ([xshift=-6pt,yshift=-10pt]n3.south west);

        \draw[aux] (edf) -- (cloud);
    \end{tikzpicture}
    \caption{AnyLog supplies the shared operational view for mobile autonomous systems, while a Byzantine-resilient agreement protocol resolves disputed state only when needed. The safety-critical path remains local; cloud, datacenter, or satellite links remain optional for summaries, history, and model distribution.}
    \label{fig:mobile-autonomy-agreement}
\end{figure}

\subsection{Enabling Federated Learning}

Federated learning allows multiple sites to collaboratively train a model without centralizing raw data~\cite{fl1,fl2,fl3}. Each participant trains on its local data and shares only model updates, which are aggregated into a new global model. This preserves data locality, ownership, and privacy while incorporating patterns learned across participating sites.

This model aligns naturally with the AnyLog Edge Data Fabric. Local databases provide access to training data, the Distributed Metadata Layer coordinates participants, policies, and training rounds, and file services exchange model artifacts. The logical query interface binds each training task to authorized local tables without moving the underlying data.

EdgeFL~\cite{edgefl} is a programming framework that uses AnyLog as the infrastructure and orchestration layer for federated learning and real-time distributed inference. It supports both designated-aggregator and decentralized configurations, allowing participating agents to coordinate training and aggregate updates independently. In either model, training rounds are coordinated through the Shared Metadata Layer. Each training agent monitors the metadata layer, queries its authorized local data through SQL, trains independently, and publishes its model update. Because AnyLog already provides distributed data access, authorization, coordination, and artifact exchange, federated learning and inference operate as native applications on the Edge Data Fabric without requiring a separate infrastructure stack.

\section{Conclusion}
\label{ch:conclusion}

\subsection{Looking Beyond the Cloud}

The future of distributed computing is not cloud-only or edge-only, but a coordinated model in which workloads execute where it is most effective. Cloud platforms remain valuable for large-scale model training, enterprise analytics, long-term storage, and fleet-wide reporting, while time-sensitive inference and operational decisions increasingly require direct access to current conditions near sensors, machines, and physical processes~\cite{satyanarayanan2017emergence,iorga2018fog}.

AnyLog enables this model by treating the cloud as one participant in a distributed computing environment rather than the operational destination for all data and computation. Through the Distributed Metadata Layer, distributed query execution, Unified Namespace, and Single System Image, the AnyLog platform enables data, services, and AI to operate as a single logical system while allowing operational data to remain at its source. Policies determine what remains local, what is processed at the edge, and which data, results, or workloads are forwarded to cloud resources.

\subsection{An Architecture Designed to Evolve}

Industrial systems will continue to evolve as new protocols, databases, storage technologies, hardware accelerators, AI models, and communication methods emerge across increasingly distributed environments. A platform tightly coupled to any one technology risks becoming obsolete or requiring costly redesign whenever that dependency changes.

AnyLog absorbs this evolution by separating the logical platform presented to users from the physical technologies that implement it. Connectors isolate data acquisition behind configurable southbound interfaces; logical tables separate applications from distributed databases; metadata decouples resource discovery from physical topology; the Unified Namespace separates asset identity from storage organization; northbound APIs hide distributed execution from consuming applications; and MCP provides AI agents with a standardized interface to the Edge Data Fabric.

Protocols, databases, models, agents, and services can as a result be introduced, upgraded, or replaced by modifying the appropriate connector, policy, metadata, or configuration layer rather than redesigning every dependent application. Although new technologies may still require integration work, AnyLog confines that work to its standardized or custom northbound and southbound interfaces while preserving a consistent Single System Image across the operational environment.

\subsection{Final Thoughts}

The central challenge at the operational edge is no longer simply collecting more data. It is making distributed data immediately understandable, accessible, and actionable without forcing every site, machine, vehicle, and application through a centralized decision path. AnyLog addresses this challenge by turning independently operating edge systems into one cooperative platform while preserving their ability to function locally.

This creates a practical foundation for industrial AI and automation. Applications and AI agents can discover available resources, interpret operational context, query current and historical data, and invoke authorized services through a consistent logical environment. Decisions execute at or near the physical systems they affect, while enterprise and cloud platforms remain available for training, long-term analysis, reporting, compliance, and other workloads suited to centralized infrastructure.

AnyLog gives organizations a practical path to seamlessly modernize operations without replacing the systems already in place. It begins with one machine, production line, facility, or vehicle fleet; connect existing data sources; expose them through the Single System Image; and expand by adding AnyLog agents as requirements and trust grow. What begins as a focused solution to today's needs becomes infrastructure that grows more capable and valuable with every new asset, site, service, and intelligent application.

\bibliographystyle{IEEEtran}
\bibliography{sample}

@IEEEtranBSTCTL{BSTcontrol,
  CTLdash_repeated_names = "no"
}

@online{ibmIEAMOverview,
  author  = {{IBM}},
  title   = {Overview of IBM Edge Application Manager},
  url     = {https://www.ibm.com/docs/en/eam/4.5.x?topic=overview-ieam},
  note    = {IBM Edge Application Manager documentation, version 4.5.x},
  urldate = {2026-07-15}
}

@misc{anylogDeploymentScripts,
  author       = {{AnyLog}},
  title        = {{AnyLog Deployment and Testing Scripts}},
  year         = {2026},
  howpublished = {\url{https://github.com/AnyLog-co/deployment-scripts}},
  note         = {GitHub repository, accessed July 15, 2026}
}

@inproceedings{abadi2020anylog,
  title={Anylog: a grand unification of the internet of things},
  author={Abadi, Daniel and Arden, Owen and Nawab, Faisal and Shadmon, Moshe},
  booktitle={Conference on Innovative Data Systems Research (CIDR ‘20)},
  year={2020}
}

@inproceedings{nawab2024tipping,
  title={The tipping point of edge-cloud data management},
  author={Nawab, Faisal and Shadmon, Moshe},
  year={2024},
  organization={CIDR}
}

@misc{shadmon2020peer,
  author       = {Moshe Shadmon and Levy Cohen and Daniel Abadi and Owen Arden},
  title        = {System and Apparatus to Manage Data Using a Peer-to-Peer Network and the Blockchain},
  howpublished = {U.S. Patent 10,868,865},
  number       = {US10868865B2},
  year         = {2020},
  month        = dec,
  day          = {15},
  url          = {https://patents.google.com/patent/US10868865B2/en},
  note         = {Filed November 20, 2018; issued December 15, 2020}
}

@article{peter2024impact,
  title={The impact of unified namespace in industry 4.0},
  author={P{\'e}ter, {\'A}d{\'a}m and Werner, Samuel},
  year={2024}
}

@article{dean2008mapreduce,
  title={MapReduce: simplified data processing on large clusters},
  author={Dean, Jeffrey and Ghemawat, Sanjay},
  journal={Communications of the ACM},
  volume={51},
  number={1},
  pages={107--113},
  year={2008},
  publisher={ACM New York, NY, USA}
}

@article{shi2016edge,
  title={Edge computing: Vision and challenges},
  author={Shi, Weisong and Cao, Jie and Zhang, Quan and Li, Youhuizi and Xu, Lanyu},
  journal={IEEE internet of things journal},
  volume={3},
  number={5},
  pages={637--646},
  year={2016},
  publisher={IEEE}
}

@article{cattell2011scalable,
  title={Scalable SQL and NoSQL data stores},
  author={Cattell, Rick},
  journal={Acm Sigmod Record},
  volume={39},
  number={4},
  pages={12--27},
  year={2011},
  publisher={ACM New York, NY, USA}
}

@article{dewitt1992parallel,
  title={Parallel database systems: The future of high performance database systems},
  author={DeWitt, David and Gray, Jim},
  journal={Communications of the ACM},
  volume={35},
  number={6},
  pages={85--98},
  year={1992},
  publisher={ACM New York, NY, USA}
}

@article{satyanarayanan2017emergence,
  title={The emergence of edge computing},
  author={Satyanarayanan, Mahadev},
  journal={computer},
  volume={50},
  number={1},
  pages={30--39},
  year={2017},
  publisher={IEEE}
}

@inproceedings{ali2021hidden,
  title={The hidden cost of the edge: a performance comparison of edge and cloud latencies},
  author={Ali-Eldin, Ahmed and Wang, Bin and Shenoy, Prashant},
  booktitle={Proceedings of the International Conference for High Performance Computing, Networking, Storage and Analysis},
  pages={1--12},
  year={2021}
}

@article{kusiak2018smart,
  title={Smart manufacturing},
  author={Kusiak, Andrew},
  journal={International journal of production Research},
  volume={56},
  number={1-2},
  pages={508--517},
  year={2018},
  publisher={Taylor \& Francis}
}

@article{wan2020artificial,
  title={Artificial-intelligence-driven customized manufacturing factory: key technologies, applications, and challenges},
  author={Wan, Jiafu and Li, Xiaomin and Dai, Hong-Ning and Kusiak, Andrew and Martinez-Garcia, Miguel and Li, Di},
  journal={Proceedings of the IEEE},
  volume={109},
  number={4},
  pages={377--398},
  year={2020},
  publisher={IEEE}
}

@article{madden2002tag,
  title={TAG: A tiny aggregation service for ad-hoc sensor networks},
  author={Madden, Samuel and Franklin, Michael J and Hellerstein, Joseph M and Hong, Wei},
  journal={ACM SIGOPS Operating Systems Review},
  volume={36},
  number={SI},
  pages={131--146},
  year={2002},
  publisher={ACM New York, NY, USA}
}

@article{khan2021blockchain,
  title={Blockchain smart contracts: Applications, challenges, and future trends},
  author={Khan, Shafaq Naheed and Loukil, Faiza and Ghedira-Guegan, Chirine and Benkhelifa, Elhadj and Bani-Hani, Anoud},
  journal={Peer-to-peer Networking and Applications},
  volume={14},
  number={5},
  pages={2901--2925},
  year={2021},
  publisher={Springer}
}

@inproceedings{zheng2024decentagram,
  title={Decentagram: Highly-Available Decentralized Publish/Subscribe Systems},
  author={Zheng, Haofan and Tran, Tuan and Shadmon, Roy and Arden, Owen},
  booktitle={2024 54th Annual IEEE/IFIP International Conference on Dependable Systems and Networks (DSN)},
  pages={274--287},
  year={2024},
  organization={IEEE}
}

@incollection{ding2005peer,
  title={Peer-to-peer networks for content sharing},
  author={Ding, Choon Hoong and Nutanong, Sarana and Buyya, Rajkumar},
  booktitle={Peer-to-Peer Computing: The Evolution of a Disruptive Technology},
  pages={28--65},
  year={2005},
  publisher={IGI Global Scientific Publishing}
}

@inproceedings{cohen2003incentives,
  title={Incentives build robustness in BitTorrent},
  author={Cohen, Bram and others},
  booktitle={Workshop on Economics of Peer-to-Peer systems},
  volume={6},
  pages={68--72},
  year={2003}
}

@book{mahnke2009opc,
  title={OPC unified architecture},
  author={Mahnke, Wolfgang and Leitner, Stefan-Helmut and Damm, Matthias},
  year={2009},
  publisher={Springer Science \& Business Media}
}

@article{mishra2020use,
  title={The use of MQTT in M2M and IoT systems: A survey},
  author={Mishra, Biswajeeban and Kertesz, Attila},
  journal={Ieee Access},
  volume={8},
  pages={201071--201086},
  year={2020},
  publisher={Ieee}
}

@article{thomas2008introduction,
  title={Introduction to the modbus protocol},
  author={Thomas, George},
  journal={The Extension},
  volume={9},
  number={4},
  pages={1--4},
  year={2008}
}

@book{murali2019hands,
  title={Hands-On RESTful API Design Patterns and Best Practices: Design, develop, and deploy highly adaptable, scalable, and secure RESTful web APIs},
  author={Murali, Anupama and Raj, Pethuru and others},
  year={2019},
  publisher={Packt Publishing Ltd}
}

@article{wang1993grpc,
  title={GRPC: A communication cooperation mechanism in distributed systems},
  author={Wang, Xingwei and Zhao, Hong and Zhu, Jiakeng},
  journal={ACM SIGOPS Operating Systems Review},
  volume={27},
  number={3},
  pages={75--86},
  year={1993},
  publisher={ACM New York, NY, USA}
}

@online{grpcDocumentation,
  author  = {{gRPC Authors}},
  title   = {Introduction to gRPC},
  year    = {2024},
  url     = {https://grpc.io/docs/what-is-grpc/introduction/},
  urldate = {2026-07-17},
  note    = {Official gRPC documentation}
}

@article{hou2025model,
  title={Model context protocol (mcp): Landscape, security threats, and future research directions},
  author={Hou, Xinyi and Zhao, Yanjie and Wang, Shenao and Wang, Haoyu},
  journal={ACM Transactions on Software Engineering and Methodology},
  year={2025},
  publisher={ACM New York, NY}
}

@article{elmenreich2002introduction,
  title={An introduction to sensor fusion},
  author={Elmenreich, Wilfried},
  journal={Vienna University of Technology, Austria},
  volume={502},
  number={1-28},
  pages={37},
  year={2002}
}

@book{iyengar2004distributed,
  title={Distributed sensor networks},
  author={Iyengar, S Sitharama and Brooks, Richard R and others},
  year={2004},
  publisher={Chapman and Hall/CRC}
}

@inproceedings{kafka,
  title = {Kafka: {{A}} Distributed Messaging System for Log Processing},
  booktitle = {Proceedings of the {{NetDB}}},
  author = {Kreps, Jay and Narkhede, Neha and Rao, Jun and others},
  year = {2011},
  volume = {11},
  pages = {1--7},
  publisher = {Athens, Greece}
}

@misc{lfedge_open_horizon,
  author       = {{LF Edge}},
  title        = {{Open Horizon}},
  howpublished = {\url{https://lfedge.org/projects/open-horizon/}},
  note         = {Accessed: July 13, 2026}
}

@misc{dell_nativeedge, author = {{Dell Technologies}}, title = {{Dell Distributed Private Cloud (formerly Dell NativeEdge)}}, howpublished = {\url{https://www.dell.com/en-us/shop/storage-servers-and-networking-for-business/sf/nativeedge}}, note = {Accessed: July 13, 2026} 
}

@article{rose2020zero,
  title={Zero trust architecture},
  author={Rose, Scott and Borchert, Oliver and Mitchell, Stu and Connelly, Sean},
  journal={NIST special publication},
  volume={800},
  number={207},
  pages={1--52},
  year={2020}
}

@misc{mcpSpecification,
  author       = {{Model Context Protocol}},
  title        = {Model Context Protocol Specification},
  year         = {2025},
  howpublished = {\url{https://modelcontextprotocol.io/specification/2025-11-25}},
  note         = {Specification version 2025-11-25; accessed July 13, 2026}
}

@article{zhou2019edge,
  title={Edge intelligence: Paving the last mile of artificial intelligence with edge computing},
  author={Zhou, Zhi and Chen, Xu and Li, En and Zeng, Liekang and Luo, Ke and Zhang, Junshan},
  journal={Proceedings of the IEEE},
  volume={107},
  number={8},
  pages={1738--1762},
  year={2019},
  publisher={IEEE}
}

@article{iorga2018fog,
  title={Fog computing conceptual model},
  author={Iorga, Michaela and Feldman, Larry and Barton, Robert and Martin, Michael J and Goren, Nedim S and Mahmoudi, Charif},
  year={2018},
  publisher={Michaela Iorga, Larry Feldman, Robert Barton, Michael J. Martin, Nedim S~…}
}

@article{standard2019mqtt,
  title={MQTT Version 5.0},
  author={Standard, OASIS},
  journal={Retrieved June},
  volume={22},
  number={2020},
  pages={1435},
  year={2019}
}

@techreport{eclipseSparkplug2022,
  author      = {{Eclipse Sparkplug Contributors}},
  title       = {Sparkplug Specification},
  institution = {Eclipse Foundation},
  type        = {Specification},
  number      = {3.0.0},
  year        = {2022},
  month       = nov,
  url         = {https://sparkplug.eclipse.org/specification/version/3.0/documents/sparkplug-specification-3.0.0.pdf},
  note        = {Accessed: July 14, 2026}
}

@inproceedings{shadmon2025enhancing,
  title={Enhancing accuracy in approximate byzantine agreement with bayesian inference},
  author={Shadmon, Roy and Arden, Owen},
  booktitle={2025 55th Annual IEEE/IFIP International Conference on Dependable Systems and Networks-Supplemental Volume (DSN-S)},
  pages={191--195},
  year={2025},
  organization={IEEE}
}

@InProceedings{shadmon2025brief,
  author =	{Shadmon, Roy and Arden, Owen},
  title =	{{Brief Announcement: Proximal Byzantine Agreement: Improved Accuracy for Fault-Tolerant Replicated Datastreams}},
  booktitle =	{39th International Symposium on Distributed Computing (DISC 2025)},
  pages =	{64:1--64:8},
  series =	{Leibniz International Proceedings in Informatics (LIPIcs)},
  ISBN =	{978-3-95977-402-4},
  ISSN =	{1868-8969},
  year =	{2025},
  volume =	{356},
  editor =	{Kowalski, Dariusz R.},
  publisher =	{Schloss Dagstuhl -- Leibniz-Zentrum f{\"u}r Informatik},
  address =	{Dagstuhl, Germany},
  URL =		{https://drops.dagstuhl.de/entities/document/10.4230/LIPIcs.DISC.2025.64},
  URN =		{urn:nbn:de:0030-drops-248808},
  doi =		{10.4230/LIPIcs.DISC.2025.64}
}

@techreport{dod2020c3modernization,
  author      = {{U.S. Department of Defense}},
  title       = {Department of Defense Command, Control, and Communications
                 ({C3}) Modernization Strategy},
  institution = {U.S. Department of Defense},
  year        = {2020},
  month       = sep,
  url         = {https://dodcio.defense.gov/Portals/0/Documents/DoD-C3-Strategy.pdf},
  note        = {Accessed: July 14, 2026}
}

@article{cai2022rbaas,
  title={RBaaS: A robust blockchain as a service paradigm in cloud-edge collaborative environment},
  author={Cai, Zhengong and Yang, Guozheng and Xu, Shaoyong and Zang, Cheng and Chen, Jiajun and Hang, Pingping and Yang, Bowei},
  journal={IEEE Access},
  volume={10},
  pages={35437--35444},
  year={2022},
  publisher={IEEE}
}

@article{lamport1983weak,
  title={The weak Byzantine generals problem},
  author={Lamport, Leslie},
  journal={Journal of the ACM (JACM)},
  volume={30},
  number={3},
  pages={668--676},
  year={1983},
  publisher={ACM New York, NY, USA}
}

@article{dolev1986reaching,
  title={Reaching approximate agreement in the presence of faults},
  author={Dolev, Danny and Lynch, Nancy A and Pinter, Shlomit S and Stark, Eugene W and Weihl, William E},
  journal={Journal of the ACM (JACM)},
  volume={33},
  number={3},
  pages={499--516},
  year={1986},
  publisher={ACM New York, NY, USA}
}

@misc{edgefl,
  author       = {Roy Shadmon},
  title        = {{EdgeFL}: Continuous Federated Learning Across Distributed Edge Nodes},
  year         = {2026},
  howpublished = {GitHub repository},
  url          = {https://github.com/royshadmon/EdgeFL},
  note         = {Accessed: July 14, 2026}
}

@article{lu2023additive,
  title={Additive Manufacturing Data Integration and Recommended Practice},
  author={Lu, Yan and Perisic, Milica and Jones, Albert},
  year={2023}
}

@inproceedings{theodoratos1997data,
  title={Data warehouse configuration},
  author={Theodoratos, Dimitri and Sellis, Timos and others},
  booktitle={VLDB},
  volume={97},
  pages={126--135},
  year={1997}
}

@article{hai2023data,
  title={Data lakes: A survey of functions and systems},
  author={Hai, Rihan and Koutras, Christos and Quix, Christoph and Jarke, Matthias},
  journal={IEEE Transactions on Knowledge and Data Engineering},
  volume={35},
  number={12},
  pages={12571--12590},
  year={2023},
  publisher={IEEE}
}

@article{turner2021human,
  title={Human in the Loop: Industry 4.0 technologies and scenarios for worker mediation of automated manufacturing},
  author={Turner, Christopher J and Ma, Ruidong and Chen, Jingyu and Oyekan, John},
  journal={IEEE access},
  volume={9},
  pages={103950--103966},
  year={2021},
  publisher={IEEE}
}

@article{liu2019edge,
  title={Edge computing for autonomous driving: Opportunities and challenges},
  author={Liu, Shaoshan and Liu, Liangkai and Tang, Jie and Yu, Bo and Wang, Yifan and Shi, Weisong},
  journal={Proceedings of the IEEE},
  volume={107},
  number={8},
  pages={1697--1716},
  year={2019},
  publisher={IEEE}
}

@article{singh2023edge,
  title={Edge AI: a survey},
  author={Singh, Raghubir and Gill, Sukhpal Singh},
  journal={Internet of Things and Cyber-Physical Systems},
  volume={3},
  pages={71--92},
  year={2023},
  publisher={Elsevier}
}

@article{meuser2024revisiting,
  title={Revisiting edge ai: Opportunities and challenges},
  author={Meuser, Tobias and Lov{\'e}n, Lauri and Bhuyan, Monowar and Patil, Shishir G and Dustdar, Schahram and Aral, Atakan and Bayhan, Suzan and Becker, Christian and De Lara, Eyal and Ding, Aaron Yi and others},
  journal={IEEE Internet Computing},
  volume={28},
  number={4},
  pages={49--59},
  year={2024},
  publisher={IEEE}
}

@article{surianarayanan2023survey,
  title={A survey on optimization techniques for edge artificial intelligence (AI)},
  author={Surianarayanan, Chellammal and Lawrence, John Jeyasekaran and Chelliah, Pethuru Raj and Prakash, Edmond and Hewage, Chaminda},
  journal={Sensors},
  volume={23},
  number={3},
  pages={1279},
  year={2023},
  publisher={MDPI}
}

@article{gill2025edge,
  title={Edge AI: A taxonomy, systematic review and future directions},
  author={Gill, Sukhpal Singh and Golec, Muhammed and Hu, Jianmin and Xu, Minxian and Du, Junhui and Wu, Huaming and Walia, Guneet Kaur and Murugesan, Subramaniam Subramanian and Ali, Babar and Kumar, Mohit and others},
  journal={Cluster Computing},
  volume={28},
  number={1},
  pages={18},
  year={2025},
  publisher={Springer}
}

@online{dodMOSA2025,
  author  = {{Office of the Under Secretary of Defense for Research and Engineering}},
  title   = {Modular Open Systems Approach},
  year    = {2025},
  url     = {https://www.cto.mil/sea/mosa/},
  urldate = {2026-07-17},
  note    = {U.S. Department of Defense}
}

@article{kore2022api,
  title={API testing using Postman tool},
  author={Kore, Pratik Prakash and Lohar, Mayuresh Jaywant and Surve, Madhusudan Tanaji and Jadhav, Snehal},
  journal={International Journal for Research in Applied Science and Engineering Technology},
  volume={10},
  number={12},
  pages={841--843},
  year={2022}
}

@misc{anylogEDM,
  author       = {{AnyLog}},
  title        = {{AnyLog Edge Data Manager}},
  year         = {2026},
  howpublished = {\url{https://github.com/AnyLog-co/Remote-GUI}},
  note         = {GitHub repository, accessed July 17, 2026}
}

@article{tao2022digital,
  title={Digital twin modeling},
  author={Tao, Fei and Xiao, Bin and Qi, Qinglin and Cheng, Jiangfeng and Ji, Ping},
  journal={Journal of Manufacturing Systems},
  volume={64},
  pages={372--389},
  year={2022},
  publisher={Elsevier}
}

@article{singh2021digital,
  title={Digital twin: Origin to future},
  author={Singh, Maulshree and Fuenmayor, Evert and Hinchy, Eoin P and Qiao, Yuansong and Murray, Niall and Devine, Declan},
  journal={Applied System Innovation},
  volume={4},
  number={2},
  pages={36},
  year={2021},
  publisher={MDPI}
}

@article{fuller2020digital,
  title={Digital twin: Enabling technologies, challenges and open research},
  author={Fuller, Aidan and Fan, Zhong and Day, Charles and Barlow, Chris},
  journal={IEEE access},
  volume={8},
  pages={108952--108971},
  year={2020},
  publisher={IEEE}
}

@inproceedings{fl1,
  title={Communication-efficient learning of deep networks from decentralized data},
  author={McMahan, Brendan and Moore, Eider and Ramage, Daniel and Hampson, Seth and y Arcas, Blaise Aguera},
  booktitle={Artificial intelligence and statistics},
  pages={1273--1282},
  year={2017},
  organization={Pmlr}
}

@misc{gartner2026agentsprawl,
  author       = {{Gartner}},
  title        = {Gartner Identifies Six Steps to Manage AI Agent Sprawl},
  year         = {2026},
  month        = apr,
  day          = {28},
  howpublished = {\url{https://www.gartner.com/en/newsroom/press-releases/2026-04-28-gartner-identifies-six-steps-to-manage-artificial-intelligence-agent-sprawl}},
  urldate      = {2026-07-28}
}

@misc{elsayed2026agentic,
  author    = {Yomna Elsayed and Cecily Jones},
  title     = {Agentic Explainability at Scale: Between Corporate Fears and XAI Needs},
  year      = {2026},
  doi       = {10.5281/zenodo.19700280},
  publisher = {Zenodo}
}

@article{ji2021survey,
  title={A survey on knowledge graphs: Representation, acquisition, and applications},
  author={Ji, Shaoxiong and Pan, Shirui and Cambria, Erik and Marttinen, Pekka and Yu, Philip S},
  journal={IEEE transactions on neural networks and learning systems},
  volume={33},
  number={2},
  pages={494--514},
  year={2021},
  publisher={IEEE}
}

@misc{schneider2026cognite,
  author       = {{Schneider Electric}},
  title        = {Schneider Electric Announces Agreement to Acquire Cognite,
                  Unlocking a New Level of Strategic Intelligence for Industrial AI},
  year         = {2026},
  month        = jun,
  day          = {30},
  howpublished = {Financial Release},
  url          = {https://www.se.com/ww/en/assets/pdf/Schneider-Electric-announces-agreement-to-acquire-Cognite},
  urldate      = {2026-07-21}
}

@article{fl2,
  title={A review of applications in federated learning},
  author={Li, Li and Fan, Yuxi and Tse, Mike and Lin, Kuo-Yi},
  journal={Computers \& Industrial Engineering},
  volume={149},
  pages={106854},
  year={2020},
  publisher={Elsevier}
}

@article{fl3,
  title={Efficient and privacy-enhanced federated learning for industrial artificial intelligence},
  author={Hao, Meng and Li, Hongwei and Luo, Xizhao and Xu, Guowen and Yang, Haomiao and Liu, Sen},
  journal={IEEE Transactions on Industrial Informatics},
  volume={16},
  number={10},
  pages={6532--6542},
  year={2019},
  publisher={IEEE}
}

@online{mitre2025validaccounts,
  author       = {{MITRE}},
  title        = {Valid Accounts: Technique T0859},
  organization = {MITRE ATT\&CK},
  year         = {2025},
  url          = {https://attack.mitre.org/techniques/T0859/},
  urldate      = {2026-07-20}
}

@incollection{tomlinson2017introduction,
  title={Introduction to the TPM},
  author={Tomlinson, Allan},
  booktitle={Smart Cards, Tokens, Security and Applications},
  pages={173--191},
  year={2017},
  publisher={Springer}
}

@article{liu2021review,
  title={Review of digital twin about concepts, technologies, and industrial applications},
  author={Liu, Mengnan and Fang, Shuiliang and Dong, Huiyue and Xu, Cunzhi},
  journal={Journal of manufacturing systems},
  volume={58},
  pages={346--361},
  year={2021},
  publisher={Elsevier}
}

@misc{aws-s3,
  author       = {{Amazon Web Services}},
  title        = {What Is Amazon S3?},
  year         = {2026},
  howpublished = {\url{https://docs.aws.amazon.com/AmazonS3/latest/userguide/Welcome.html}},
  note         = {Amazon Simple Storage Service User Guide. Accessed: 2026-07-30}
}

@misc{mongodb-gridfs,
  author       = {{MongoDB}},
  title        = {{GridFS} for Self-Managed Deployments},
  year         = {2026},
  howpublished = {\url{https://www.mongodb.com/docs/manual/core/gridfs/}},
  note         = {MongoDB Database Manual. Accessed: 2026-07-30}
}

@misc{minio-object-storage,
  author       = {{MinIO}},
  title        = {MinIO Object Storage for Linux},
  year         = {2025},
  howpublished = {\url{https://min.io/docs/minio/linux/index.html}},
  note         = {MinIO Documentation. Accessed: 2026-07-30}
}

@misc{aws-kinesis-video,
  author       = {{Amazon Web Services}},
  title        = {What Is Amazon Kinesis Video Streams?},
  year         = {2026},
  howpublished = {\url{https://docs.aws.amazon.com/kinesisvideostreams/latest/dg/what-is-kinesis-video.html}},
  note         = {Amazon Kinesis Video Streams Developer Guide. Accessed: 2026-07-30}
}

@misc{aws-rekognition-streaming,
  author       = {{Amazon Web Services}},
  title        = {Working with Streaming Video Events},
  year         = {2026},
  howpublished = {\url{https://docs.aws.amazon.com/rekognition/latest/dg/streaming-video.html}},
  note         = {Amazon Rekognition Developer Guide. Accessed: 2026-07-30}
}

\end{document}